\documentclass[letterpaper]{article} 
\usepackage{aaai23}  
\usepackage{times}  
\usepackage{helvet}  
\usepackage{courier}  
\usepackage[hyphens]{url}  
\usepackage{graphicx} 
\usepackage{natbib}  
\usepackage{caption} 
\usepackage{algorithm}
\usepackage{algorithmic}
\usepackage{amsmath}
\usepackage{amsfonts} 
\usepackage{amssymb}
\usepackage{color}
\usepackage{cases}
\usepackage{graphicx}
\usepackage{footmisc}
\usepackage{cleveref}
\usepackage{float}

\newtheorem{thm}{Theorem}

\usepackage{newfloat}
\usepackage{listings}
\usepackage{subfigure}
\DeclareCaptionStyle{ruled}{labelfont=normalfont,labelsep=colon,strut=off} 
\floatstyle{ruled}
\newfloat{listing}{tb}{lst}{}
\floatname{listing}{Listing}
\title{RLEKF: An Optimizer for Deep Potential with \textit{Ab Initio} Accuracy}
\author{
    Siyu Hu\textsuperscript{\rm 1,2}\equalcontrib,
    Wentao Zhang\textsuperscript{\rm 3}\equalcontrib,
    Qiuchen Sha\textsuperscript{\rm 1,2},
    Feng Pan\textsuperscript{\rm 3},\\
    Lin-Wang Wang\textsuperscript{\rm 4},
    Weile Jia\textsuperscript{\rm 1},
    Guangming Tan\textsuperscript{\rm 1},
    Tong Zhao\textsuperscript{\rm 1}\thanks{Corresponding author}
}
\affiliations{
    \textsuperscript{\rm 1} State Key Lab of Processors, Institute of Computing Technology, Chinese Academy of Sciences, Beijing, China\\
    \textsuperscript{\rm 2}University of Chinese Academy of Sciences, Beijing, China\\
    \textsuperscript{\rm 3}School of Advanced Materials, Shenzhen Graduate School, Peking University, Shenzhen, China\\
    \textsuperscript{\rm 4}Institute of Semiconductors, Chinese Academy of Sciences, Beijing, China\\
    husiyu20b@ict.ac.cn, zhang\_wt@pku.edu.cn, shaqiuchen22@mails.ucas.ac.cn, panfeng@pkusz.edu.cn,\\ lwwang@semi.ac.cn, \{jiaweile,tgm,zhaotong\}@ict.ac.cn
}

\usepackage{bibentry}

\begin{document}

\newcommand{\zt}[1]{\textcolor{red}{[#1]}}
\newcommand{\WL}[1]{\textcolor{blue}{[Weile:#1]}}
\newcommand{\SY}[1]{\textcolor{cyan}{[Siyu:#1]}}

\maketitle

\begin{abstract}
	It is imperative to accelerate the training of neural network force field such as Deep Potential, which usually requires thousands of images based on first-principles calculation and a couple of days to generate an accurate potential energy surface. To this end, we propose a novel optimizer named reorganized layer extended Kalman filtering (RLEKF), an optimized version of global extended Kalman filtering (GEKF) with a strategy of splitting big and gathering small layers to overcome the $O(N^2)$ computational cost of GEKF. This strategy provides an approximation of the dense weights error covariance matrix with a sparse diagonal block matrix for GEKF. We implement both RLEKF and the baseline Adam in our $\alpha$Dynamics package and numerical experiments are performed on 13 unbiased datasets. Overall, RLEKF converges faster with slightly better accuracy. For example, a test on a typical system, bulk copper, shows that RLEKF converges faster by both the number of training epochs ($\times$11.67) and wall-clock time ($\times$1.19). Besides, we theoretically prove that the updates of weights converge and thus are against the gradient exploding problem. Experimental results verify that RLEKF is not sensitive to the initialization of weights. The RLEKF sheds light on other AI-for-science applications where training a large neural network (with tons of thousands parameters) is a bottleneck.
\end{abstract}

\section{Introduction}

\textit{Ab initio} molecular dynamics (AIMD) has been the method of choice in modeling physical phenomena from a microscopic scale, for example, water~\cite{rahman1971molecular, stillinger1974improved}, alloy~\cite{wang2009experimental},  nanotube~\cite{raty2005growth}, and even protein~\cite{karplus2005molecular}. However, the cubic scaling of the first-principles methods~\cite{meier2014solid} has hindered both spatial and temporal scales of AIMD packages within thousands of atoms and picoseconds on modern supercomputers. To overcome the ``scaling wall'' of AIMD, two types of machine-learned MD (MLMD) methods are adopted. The first one is based on classical ML methods. In 1992, Ercolessi and Adam first introduced ML for describing potentials with an accuracy comparable to that obtained by \textit{ab initio} methods~\cite{ercolessi1992interatomic,ercolessi1994interatomic}, and to date, methods like ACE~\cite{PhysRevB.99.014104, Lysogorskiy2021}, SNAP~\cite{THOMPSON2015316}, and GPR~\cite{bartok2010gaussian} are developed and widely used in physical problems such as copper and silicon, tantalum, bulk crystals. The second approach is based on neural network (NN) and was first introduced in 2007 by Belher and Parrinello~\cite{PhysRevLett.98.146401}. The neural network MD (NNMD) method approximates both atomic energy ($E_i$) and force ($F_i$) with a local neighboring environment, and the atomic potential energy surface is trained through tons of data generated from first-principles calculations. One current state-of-the-art is the Deep Potential (DP) model, which combines large NN and physical symmetries (translation, rotation, and permutation invariance) for accurately describing the high-dimensional configuration space of interatomic potential. 
Although the corresponding package DeePMD-kit can reach 10 billion atoms when scaling to the top supercomputers~\cite{9355242, 10.1145/3503221.3508425} in model inference
, training procedure of an individual model can still take from hours to days and is the bottleneck. 




The two most commonly used training methods are Adam~\cite{kingma2014adam} and scholastic gradient descent (SGD)~\cite{saad1998online} in NNMD packages due to their integration in NN framework such as TensorFlow and PyTorch. For example, many NNMD packages such as HDNNP~\cite{Behler_2014}, SIMPLE-NN~\cite{lee2019simple} adopt Adam in the training of  interatomic potential. Yet these optimizers have a slow convergence rate in searching for the optimal  solution on the landscape and can take up to hundreds of epochs in training one NNMD model with thousands of training data. Moreover, SGD suffers from gradient exploding without proper control of the learning rate.

Global extended Kalman filtering \cite{chui2017kalman} (GEKF) is a good choice in both convergence and robustness. For example, RuNNer \cite{doi:10.1063/1.3553717} adopts GEKF as an optimizer in its training of a simple three-layer fully connected NN with 1000 parameters or so. As shown in~\cite{singraber2019parallel}, RuNNer can achieve 0.69 meV/atom in Energy RMSE and 35.5 meV/$\mathring{\text{A}}$ in Force RMSE for H$_\text{2}$O  physical system. However, since the error covariance matrix $\mathbf{P}$ is updated globally (Fig.~\ref{figss}), GEKF can be computationally expensive when NNs with tens of thousands of parameters are applied.

Our \textbf{main contribution} is a reorganized layer extended Kalman filtering (RLEKF) method, which approximates the dense weights error covariance matrix of GEKF with a sparse diagonal block matrix to reduce the computational cost. Technically, these layers are reorganized by splitting big and gathering adjacent small layers. To have a fair comparison, both RLEKF and Adam methods are implemented in our NNMD package $\alpha$Dynamics.
Compared to the Adam method, our testing results show that RLEKF can reach the same or higher accuracy for both force (better than Adam in 7 out 13 testing cases) and energy (better than Adam in 11 out of 13 testing cases). For a typical copper system, the time-to-solution of RLEKF can be $\times11.67$ and $\times1.19$ faster in terms of the number of epochs and wall-clock time, respectively. Especially, RLEKF can significantly reduce the number of epochs to 2-3 for reaching a reasonable accuracy ($1.2\times$ the RMSE of the best accuracy possible).
We theoretically prove the weights updating convergence and therefore this protects RLEKF from gradient exploding. Our work also sheds light on other NN-based applications where training NNs with a relatively large number of parameters is a bottleneck.

\begin{figure*}[t]
	\centering
	\includegraphics[width=\textwidth]{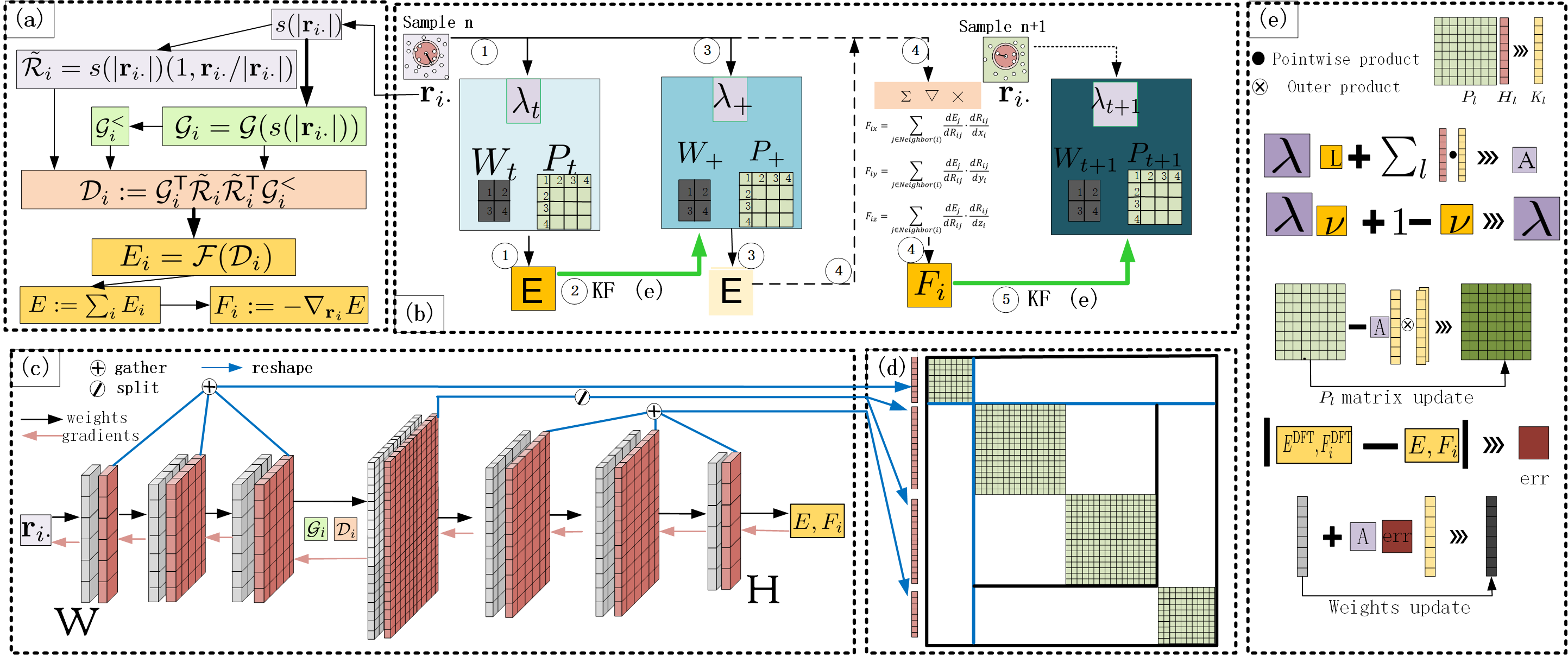} 
	\caption{The overview of DP NN with RLEKF. (a) Structure of DP NN. The two arrows in bold correspond to the embedding net and the fitting net. (b) Macro-structure of training,  $\textcircled{1}$ input data into the network, $\textcircled{2}$ calculate the gradient of energy with respect to weights through backpropagation and update $\boldsymbol{w}, \mathbf{P}, \lambda$ with Kalman filter $\textcircled{3}$ feed data into the network for energy as an intermediate, $\textcircled{4}$ obtain force through the energy obtained in $\textcircled{3}$ based on a formula the arrow directs to, $\textcircled{5}$ derive the gradient of force with respect to weights through backpropagation and use Kalman filtering to update $\boldsymbol{w}, \mathbf{P}, \lambda$, and then the above progress for the next sample starts. (c) Weights and gradients flow of DP NN. The \textbf{gather} OP is implemented on layers when the total number of accumulated parameters is less than $N_{b}$, whereas the \textbf{split} OP is activated on a layer if the number of its parameters exceeds $N_{b}$. (The \textbf{gather} works on the first three layers in embedding net and the last two layers in fitting net, meanwhile the \textbf{split} works on the first layer in fitting net in the following experiment.) (d) Error covariance matrix and splitting strategy. This picture shows our splitting strategy which is applied to most of the following experiments. 
		Here is an example: with the default block size set as $N_b=$5120 and 10240 for efficiency, except for the first layer of the fitting net, the number of weights of each layer, less than the default block size, so does not need to be split. The first fitting net layer containing 20000 weights will be split into 
		2 parts ($20000= 1 \times 10240+ 1 \times 9760 $) if $N_{b}=10240$. 
		(e) Updating strategy (Alg.~\ref{alg2}).}
	\label{figovv}
\end{figure*}
\section{Related Work}
\textbf{NNMD Packages.}
Fully connected NNs are the most widely used in NNMD packages. For example, HDNNP~\cite{PhysRevLett.98.146401, behler2017first}, which is a three-layered fully-connected NN is introduced in RuNNer~\cite{doi:10.1063/1.3553717}. Other packages such as BIM-NN~\cite{doi:10.1021/acs.jpclett.7b01072}, Simple-NN \cite{lee2019simple}, CabanaMD-NNP \cite{DESAI2022108156}, SPONGE \cite{https://doi.org/10.1002/cjoc.202100456}, DeepMD-kit~\cite{wang2018deepmd}
are also implemented via fully-connected NN. Many physical phenomena such as a diagram of water and Cu$_\text{2}$S ~\cite{singraber2019parallel},  
chemical molecule~\cite{doi:10.1021/acs.jpclett.7b01072}, 
SiO$_\text{2}$~\cite{lee2019simple}, 
organic molecules (C$_\text{10}$H$_\text{2}$, C$_\text{10}$H$_\text{3}^+$)~\cite{Lysogorskiy2021} 
are studied with the packages above.

Recent progress of NNMD packages implemented with graph NN (GNN) is gaining momentum. DimeNet++~\cite{gasteiger_dimenet_2020, gasteiger_dimenetpp_2020}, NequIP~\cite{Batzner2022}, has shown great potential in describing organic molecules (QM7 and QM9 dataset) at high accuracy. 
We remark that NNMD packages also differ in employing physical symmetries for NN inputs (``features''), which are not discussed due to it is out of the scope of this paper.  

\noindent\textbf{Training Methods in NNMD Packages.} Nearly all NNMD packages mentioned above use Adam and SGD as the training procedure for their ease of use in NN frameworks. One challenge of these methods is their time-consuming model training. For example, it usually takes more than 100 epochs to systematically train an individual DP model in the DeePMD-kit software. 

As an alternative, Kalman filtering~\cite{kalman1960contributions, welch1995introduction} (KF) aims to estimate states of a linear process theoretically based on a state-measurement model with Gaussian noise by an estimator optimal in the sense of minimizing the mean square error of predicted states with the noisy measurement as input. {It is widely used in  autonomous, navigation, and interactive computer graphics due to its fast convergence and noise filtering.} 
However, the formulation of KF confines itself to a linear optimization problem. 
Extended Kalman filtering~\cite{smith1962application} (EKF) is introduced to solve nonlinear optimization problems through Taylor expansion, and then the linearized problem is solved by KF. In the implementation, EKF has many variants such as NDEKF \cite{367790}, ONDEKF, LDEKF, and FDEKF. Note that the performance of EKF and its variants are compared in Ref.~\cite{728124}.

We remark that in training an NNMD model with more than tens of thousands of parameters, such as the DP mentioned in this paper, both Adam and EKF (and its currently known variants) are either not effective or not efficient. 

\section{Problem Setup and Formulation}

\noindent\textbf{DP.} The key steps of the DP training are shown in Fig.~\ref{figovv}(a). For each atom $i$, the physical symmetries such as  translational, rotational, and permutational invariances are integrated into the descriptor $\mathcal{D}_{i}$ through a three-layer fully connected network named embedding net. Then $\mathcal{D}_{i}$ is trained via a three-layered fitting net. 
\begin{enumerate}
	\item Every snapshot of the molecule system consists of each atom's  3D Cartesian coordinates $\mathbf{r}_{i}=(x_{i},y_{i},z_{i})\in R^{3}, i=1,2,...,N_{a}$ which is then translated into  neighbor list of atom $i$, $\mathcal{R}_{i}=\{\mathbf{r}_{ij}\in \mathbb{R}^3|\mid \mathbf{r}_{j}-\mathbf{r}_{i}\mid <r_{c}\}$ as the input of DP NN. Then, we gather it into its smooth version $\tilde{\mathcal{R}}_{i} \in \mathbb{R}^{N_{m}\times4}$, $(\tilde{\mathcal{R}}_{i})_{j}=s(|\mathbf{r}_{ij}|)(1,\mathbf{r}_{ij}/|\mathbf{r}_{ij}|)\in \mathbb{R}^4$, where $s(x)=1/x$ when $x<r_{cs}$, $s(x)=0$ when $x>r_{c}$, decaying smoothly between the two thresholds, $N_{m}$ is the maximum length of all neighbor lists, $j$ is the neighbor index of atom $i$. 
	\item Define the embedding net $\mathcal{G}_{i}\in \mathbb{R}^{N_{m}\times M}$, $\mathcal{G}_{i}=\mathcal{G}(s(|\mathbf{r}_{i\cdot}|))$, where $M$ is called symmetry order, $\mathcal{G}=\mathcal{E}_{2}\circ\mathcal{E}_{1}\circ\mathcal{E}_{0}$, $\mathcal{E}_{l}(X)=X+\tanh \left(XW_{l}+\boldsymbol{1}\otimes w_{l}\right), l \in\{1, 2\}$, $\mathcal{E}_{0}(\mathbf{x})=\tanh \left(\mathbf{x} \otimes W_{0}+\boldsymbol{1}\otimes w_{0}\right)$, $|\mathbf{r}_{i\cdot}| $ and $ \boldsymbol{1}\in \mathbb{R}^{N_{m}\times1}, w_{l} \in \mathbb{R}^{1 \times M}, W_{l}\in \mathbb{R}^{N_{m} \times M}, l \in\{0,1,2\}$ and the functions $s$ and $\tanh$ are element-wise. 
	
	\item $\tilde{\mathcal{R}}_{i}$ timing $\mathcal{G}_{i}$ yields the so-called descriptor $\mathcal{D}_{i}:=\mathcal{G}_{i}^{\textsf{T}}\tilde{\mathcal{R}}_{i}\tilde{\mathcal{R}}_{i}^{\textsf{T}}\mathcal{G}_{i}^{<}\in \mathbb{R}^{M\times M^{<}}$, i.e. $\mathcal{G}_{i}^{<}$ is the several columns of $\mathcal{G}_{i}$ and $M^{<}<M$. 
	
	\item The fitting net is  $E_{i}=\mathcal{F}(\mathcal{D}_{i})=\mathcal{F}_{3}\circ\mathcal{F}_{2}\circ\mathcal{F}_{1}\circ\mathcal{F}_{0}(\mathcal{D}_{i})$, where $\mathcal{D}_{i}$ is reshaped into a vector of form $\mathbb{R}^{MM^{<}\times 1}$, $\mathcal{F}_{l}(\mathbf{x})=\mathbf{x}+\tanh \left(\tilde{W}_{l}\mathbf{x}+\tilde{w}_{l}\right), \tilde{w}_{l} \in \mathbb{R}^{d \times 1}, \tilde{W}_{l}\in \mathbb{R}^{d \times d},l \in\{1, 2\}$, $\mathcal{F}_{3}(\mathbf{x})=\tilde{W}_{3}\mathbf{x}+\tilde{w}_{3}$, $\tilde{w}_{3} \in \mathbb{R}$,  $\tilde{W}_{3}\in \mathbb{R}^{1 \times d}$, $\mathcal{F}_{0}(\mathbf{x})=\tanh \left(\tilde{W}_{0}\mathbf{x}+\tilde{w}_{0}\right)$ , $\tilde{w}_{0} \in \mathbb{R}^{d \times 1}$, $\tilde{W}_{0}\in \mathbb{R}^{d \times MM^{<}}$. 
	\item The output $E:=\sum_{i}E_{i}$, $F_{i}:=-\nabla_{\mathbf{r}_{i}}E$.
\end{enumerate}

\noindent\textbf{EKF with Memory Factor for NNs.}
For neural networks, the model of interest 
\begin{equation}
	\begin{cases}
		\boldsymbol{\theta}_{t}=\boldsymbol{\theta}_{t-1}=\boldsymbol{w},\\
		y_{t}=h(\boldsymbol{\theta}_{t},x_{t})+\eta_{t}, \quad \eta_{t}\sim{\mathcal {N}} (0, R_{t}),
	\end{cases}
	\label{eq:Taylor approximation}
\end{equation}
is formulated in stochastic language as an EKF problem targeting on $\boldsymbol{\hat{\theta}}_{t}$, where $\boldsymbol{w}$ is the vector of all trainable parameters in the network $h(\cdot, \cdot)$,  $\{(x_{t},y_{t})\}_{t\in \mathbb{N}}$ are pairs of feature and label, $\{y_{t}\}_{t\in{\mathbb{N}}}$ can also be seen as measurements of EKF, $\{\eta_{t}\}_{t\in\mathbb{N}}$ are noise terms subject to normal distribution with mean $0$ and variances $\{R_{t}\}_{t\in\mathbb{N}}$ correspondingly, and $\forall t \in \mathbb{N}, \boldsymbol{\hat{\theta}}_{t|t-1}:=\mathbb{E}[\boldsymbol{\theta}_{t}|y_{t-1},y_{k-2},\dots,y_{1}]$. With fixed $\boldsymbol{\theta}_{t}$ and bounded $x_{t}$, $h(\boldsymbol{\theta}_{t},x_{t})$ will be approximated well by its linearization at $\boldsymbol{\hat{\theta}}_{t|t-1}$, just omitting a term $\mathcal{O}((\boldsymbol{\theta}_{t}-\boldsymbol{\hat{\theta}}_{t|t-1})^{2})$.
\begin{align}
	y_{t}&\approx h(\boldsymbol{\hat{\theta}}_{t|t-1},x_{t})+\mathbf {H} _{t}^{\textsf {T}}(\boldsymbol{\theta}_{t}-\boldsymbol{\hat{\theta}}_{t|t-1})+\eta_{t}\label{eq:Taylor}\\
	\mathbf {H} _{t}&=\frac{\partial h(\boldsymbol{\theta},x_{t})}{\partial \boldsymbol{\theta}}\bigg|_{ \boldsymbol{\theta}= \boldsymbol{\hat{\theta}}_{t|t-1}} \nonumber
\end{align}
If set $m_{t}=y_{t}- h(\boldsymbol{\hat{\theta}}_{t|t-1},x_{t})+\mathbf {H} _{t}^{\textsf {T}}\boldsymbol{\hat{\theta}}_{t|t-1}$ and rewrite \eqref{eq:Taylor} the following KF problem
\begin{numcases}{}
	\boldsymbol{\theta}_{t}=\boldsymbol{\theta}_{t-1}=\boldsymbol{w},\\
	m_{t}\approx \mathbf {H} _{t}^{\textsf {T}}\boldsymbol{\theta}_{t}+\eta_{t} .
\end{numcases}
At the beginning of training, the estimator $\boldsymbol{\hat{\theta}}_{t|t-1}$ is far away from $\boldsymbol{w}$,  so less attention should be paid to those data fed to the network at an earlier stage of training than those at later stage.  Through timing a factor $\alpha_{t}:=\Pi_{i=1}^{t} \lambda_{i}^{-1/2}$ and $\alpha_{0}:=1$, where $0< \lambda_{i}\leq 1$ and $\lambda_{i} \rightarrow 1$, the last problem enjoys the better variant as below
\begin{equation}
	\begin{cases}
		\boldsymbol{\theta}_{t}=\lambda_{t}^{-1/2}\boldsymbol{\theta}_{t-1}, \	\boldsymbol{\theta}_{1}=\boldsymbol{w}\\
		\tilde{m}_{t}=\alpha_{t}m_{t}\approx  \mathbf {H} _{t}^{\textsf {T}}\boldsymbol{\theta}_{t}+\alpha_{t}\eta_{t} =\mathbf {H} _{t}^{\textsf {T}}\boldsymbol{\theta}_{t}+\tilde{\eta}_{t}, 
	\end{cases}
	\label{eq:ekf}
\end{equation}
where $\lambda_{t}$ is called memory factor. The greater  $\lambda_{t}$ is, the more weight, or say attention, is paid to previous data.
According to basic KF theory~\cite{haykin2001kalman} , we obtain
\begin{equation*}
	\begin{split}
		\mathbf {a} _{t}&=\lambda _{t}^{-1}\mathbf {H}^{\textsf {T}} _{t}\mathbf {P} _{t-1}\mathbf {H} _{t}+\alpha_{t}^{2}R _{t},\\
		\mathbf {K} _{t}&=\lambda _{t}^{-1}\mathbf {P} _{t-1}\mathbf {H} _{t}^{\textsf {T}}\mathbf {a} _{t}^{-1},\\
		\mathbf {P} _{t}&=\left(\mathbf {I} -\mathbf {K} _{t}\mathbf {H} _{t}\right)\lambda _{t}^{-1}\mathbf {P} _{t-1},\\
		\hat {\boldsymbol{\theta}}_{t}&={\hat {\boldsymbol{\theta} }}_{t\mid t-1}+\mathbf {K} _{t}{\tilde {\epsilon }}_{t},\\
		\tilde {\epsilon}_{t}&=\tilde{m} _{t}-\mathbf {H} _{t}^{\textsf{T}}{\hat {\boldsymbol {\theta} }}_{t\mid t-1}=\alpha_{t}(y_{t}- h(\alpha_{t}^{-1}\boldsymbol{\hat{\theta}}_{t|t-1},x_{t})).
	\end{split}
\end{equation*}
Finally, we recover the estimator of $w$ via that of $\theta$ divided by the factor  $\alpha_{t}$, define $\forall t \in \mathbb{N}, \hat {\boldsymbol{w}}_{t\mid t-1}:=\alpha_{t}^{-1}\hat {\boldsymbol{\theta}}_{t\mid t-1}, \boldsymbol{w}_{t}:=\alpha_{t}^{-1}\hat {\boldsymbol{\theta}}_{t}$, find $\hat{\boldsymbol{w}}_{t\mid t-1}=\boldsymbol{w}_{t-1}$, and then get our weights updating strategy
\begin{equation*}
	\begin{split}
		\epsilon_{t}&=y_{t}- h(\boldsymbol{w}_{t-1},x_{t}),\\
		\boldsymbol{w}_{t}&={\boldsymbol{w} }_{t-1}+\mathbf {K} _{t}{\epsilon}_{t}.
	\end{split}
\end{equation*}
\begin{table*}[htbp]
	\begin{tabular}{c|c|c|c|c|c|c|c}
		Systems & Structure & Time Step (fs)& \# Snapshots & Energy(trn) & Energy(tst) & Force(trn) &Force(tst)\\\hline  
		Cu &FCC& 1 & 1646 &\textbf{0.250}/0.451&\textbf{0.327}/0.442&40.6/\textbf{39.7}&45.2/\textbf{44.4}\\
		Si & DC  & 2.5-3.5&3000& \textbf{0.148}/0.165&0.186/\textbf{0.181}&22.3/\textbf{21.9}&24.1/\textbf{23.4}\\
		C & Graphene &2-3.5& 4000 &\textbf{0.0856}/0.133&\textbf{0.267}/0.278&\textbf{24.2}/27.6&\textbf{34.7}/35.5\\
		Ag & FCC  & 2.5-3& 2015 &\textbf{0.142}/0.159&\textbf{0.243}/0.265&11.3/\textbf{11.0}&\textbf{12.4}/12.5\\
		Mg & HCP & 0.5-2 & 4000 &\textbf{0.111}/0.160&\textbf{0.169}/0.189&\textbf{13.0}/14.8&\textbf{16.1}/17.1\\
		NaCl & FCC& 2-3.5 & 3193 &\textbf{0.0403}/0.0631&\textbf{0.0435}/0.0514&6.06/\textbf{5.78}&6.69/\textbf{6.47}\\
		Li & BCC,FCC,HCP & 0.5 & 2494 &\textbf{0.0482}/0.126&\textbf{0.312}/0.332&22.9/\textbf{14.0}&\textbf{24.8}/26.4\\
		Al & FCC & 2-3.5 & 4000 &\textbf{0.359}/0.534&\textbf{0.473}/1.24&\textbf{54.1}/56.3&\textbf{58.6}/55.8\\
		MgAlCu & Alloy &3 & 2530 &\textbf{0.165}/0.227&\textbf{0.218}/0.233&44.1/\textbf{37.6}&45.2/\textbf{42.6}\\
		H$_2$O &Liquid& 0.5 & 4000 &\textbf{0.297}/0.584&\textbf{0.545}/1.00&60.8/\textbf{20.6}&\textbf{68.1}/87.6\\
		S &  S$_8$ & 3-5 & 7000 &\textbf{0.210}/0.401&0.628/0.628&51.5/\textbf{45.9}&60.4/\textbf{58.5}\\
		CuO & FCC & 3 & 1000 & \textbf{2.203}/2.47 & \textbf{2.04}/3.78 & 424/\textbf{414} & 442/\textbf{438}\\
		Cu+C & SA &3-3.2 & 2000 &\textbf{0.458}/1.27&\textbf{1.07}/2.52&176/\textbf{143}&\textbf{190}/195\\
	\end{tabular}
	\caption{\label{tab:rmse}  Structures, the quantities of snapshots, time steps (frequency of yielding snapshots),  and the root mean square errors (RMSE) of the training and the testing of RLEKF (before slashes) and Adam (after slashes) in terms of energy (meV)  and forces (meV/$\mathring{\text{A}}$) after 30 epochs for various systems. The RMSEs of the energies are normalized by the number of atoms in the system. For all sub-systems, the former 80\% and the latter 20\% of the snapshot dataset are used for training and testing, respectively. Better results are in bold. Acronyms: FCC (face-centered cubic), BCC (body-centered cubic), HCP (hexagonal close-packed), DC (diamond cubic), SA (surface adsorption).}
\end{table*}
\begin{algorithm}
	\renewcommand{\algorithmicrequire}{\textbf{Input:}}
	\renewcommand{\algorithmicensure}{\textbf{Output:}}
	\caption{High-level Structure of Training with RLEKF}
	\label{alg1}
	\begin{algorithmic}[1]
		\REQUIRE $ \left\{ {\boldsymbol{w}_{0}} \right\},\left\{ {\mathbf{P}_{0}} \right\}, \left\{ {\lambda_{1}} \right\} $
		\FOR{$t =0, 1, 2,\dots,T-1$}
		\STATE $ \hat{E}  = h_{E}(\boldsymbol{w}_{t}, x_{t})$ \label{alg1.1}
		\STATE $ \boldsymbol{w}_{+}, \mathbf{P}_{+}, \lambda_{+} =\text{RLEKF}(\hat{E}, E_{DFT}, \boldsymbol{w}_{t}, \mathbf{P}_{t}, \lambda_{t}) $\label{alg1.2}
		\STATE $ \hat{F} = h_{F}(\boldsymbol{w}_{+})$\label{alg1.3}
		\STATE $ \boldsymbol{w}_{t+1},\!\mathbf{P}_{t+1},\!\lambda_{t+1}\!=\!\text{RLEKF}(\hat{F}, F_{DFT}, \boldsymbol{w}_{+}, \mathbf{P}_{+}, \lambda_{+})$\label{alg1.4}
		\ENDFOR
		\ENSURE $ \left\{ {\boldsymbol{w}_{T}} \right\},\left\{ {\mathbf{P}_{T}} \right\}, \left\{ {\lambda_{T}} \right\} $
	\end{algorithmic}  
\end{algorithm}
\begin{algorithm}
	\renewcommand{\algorithmicrequire}{\textbf{Input:}}
	\renewcommand{\algorithmicensure}{\textbf{Output:}}
	\caption{RLEKF$(\hat{Y}, Y^{\text{DFT}}, \boldsymbol{w}^{in}, \mathbf{P}^{in}, \lambda^{in})$}
	\label{alg2}
	\begin{algorithmic}[1]
		\FOR{$i =1,2,\dots, length(Y^{\text{DFT}})$} \label{alg2.1}
		\IF{$\hat{Y}_{i} \geq Y_{i}^{\text{DFT}}$}
		\STATE $\hat{Y}_{i} $ = -$\hat{Y}_{i}$ 
		\ENDIF
		\ENDFOR
		\STATE $ \hat{Y} =\frac{\sum_{i}\hat{Y}_{i}}{length(Y^{\text{DFT}})},  errY =\frac{\sum_{i}|Y^{\text{DFT}}_{i}-\hat{Y}_{i}|}{length(Y^{\text{DFT}})}$
		\STATE $ H  =\nabla_{w}\hat{Y}\mid_{w=\boldsymbol{w}^{in}}$  \label{alg2.2}
		\STATE $ a = 1/(L\lambda^{in} + H^{\textsf{T}} \mathbf{P}^{in} H)$  \label{alg2.3}
		\FOR{$l =1,2,\dots, L$}  
		\STATE $ K = \text{split}( \mathbf{P}^{in},l)\times \text{split}(H,l)$  \label{alg2.5}
		\STATE $ \mathbf{P}^{out}_{l}= (1/ \lambda^{in}) \times ( \text{split}( \mathbf{P}^{in},l) -aKK^{\textsf{T}} ) $  \label{alg2.6}
		\STATE $ \boldsymbol{w}^{out}_{l} = \text{split}( \boldsymbol{w}^{in} ,l)+ aK errY $ 
		\ENDFOR
		\STATE $ \lambda^{out} = \lambda^{in} \nu +1- \nu $  \label{alg2.4}
		\STATE $ \boldsymbol{w}^{out} = \text{collect}(\boldsymbol{w}^{out}_l|l=1, \dots, L) $
		\STATE $ \mathbf{P}^{out} =\text{collect}( \mathbf{P}^{out}_l|l=1, \dots, L )$
		\ENSURE $ \boldsymbol{w}^{out}, \mathbf{P}^{out}, \lambda^{out} $ \label{alg2.7}
	\end{algorithmic}
\end{algorithm}
\begin{figure}[t]
	\centering
	\includegraphics[width=0.45\textwidth]{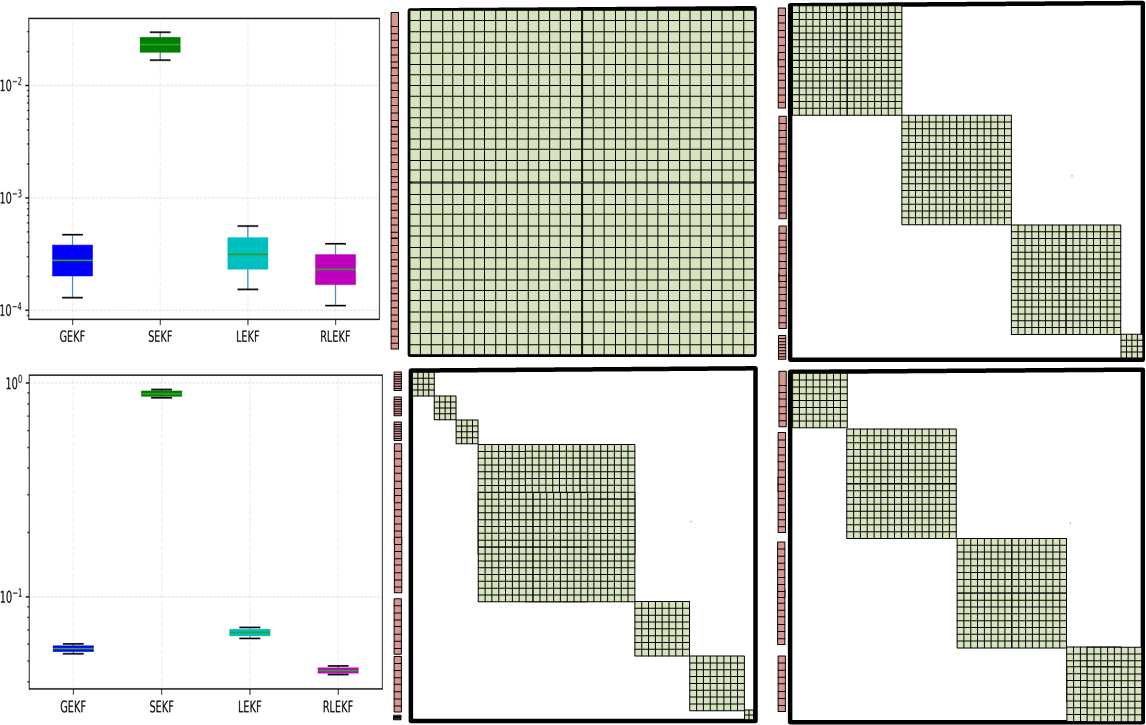}
	\caption{The two subfigures on the left are Energy (eV) (above) and Force (eV/$\mathring{\text{A}}$) (below) RMSE boxplots of test dataset under different weights splitting strategy. The four subfigures on the right are weights error covariance matrixs corresponding to GEKF, SEKF, LEKF, RLEKF in text order.}
	\label{figss}
\end{figure}
\begin{figure*}[h]
	\centering
	\subfigure{
		\includegraphics[width=0.23\textwidth]{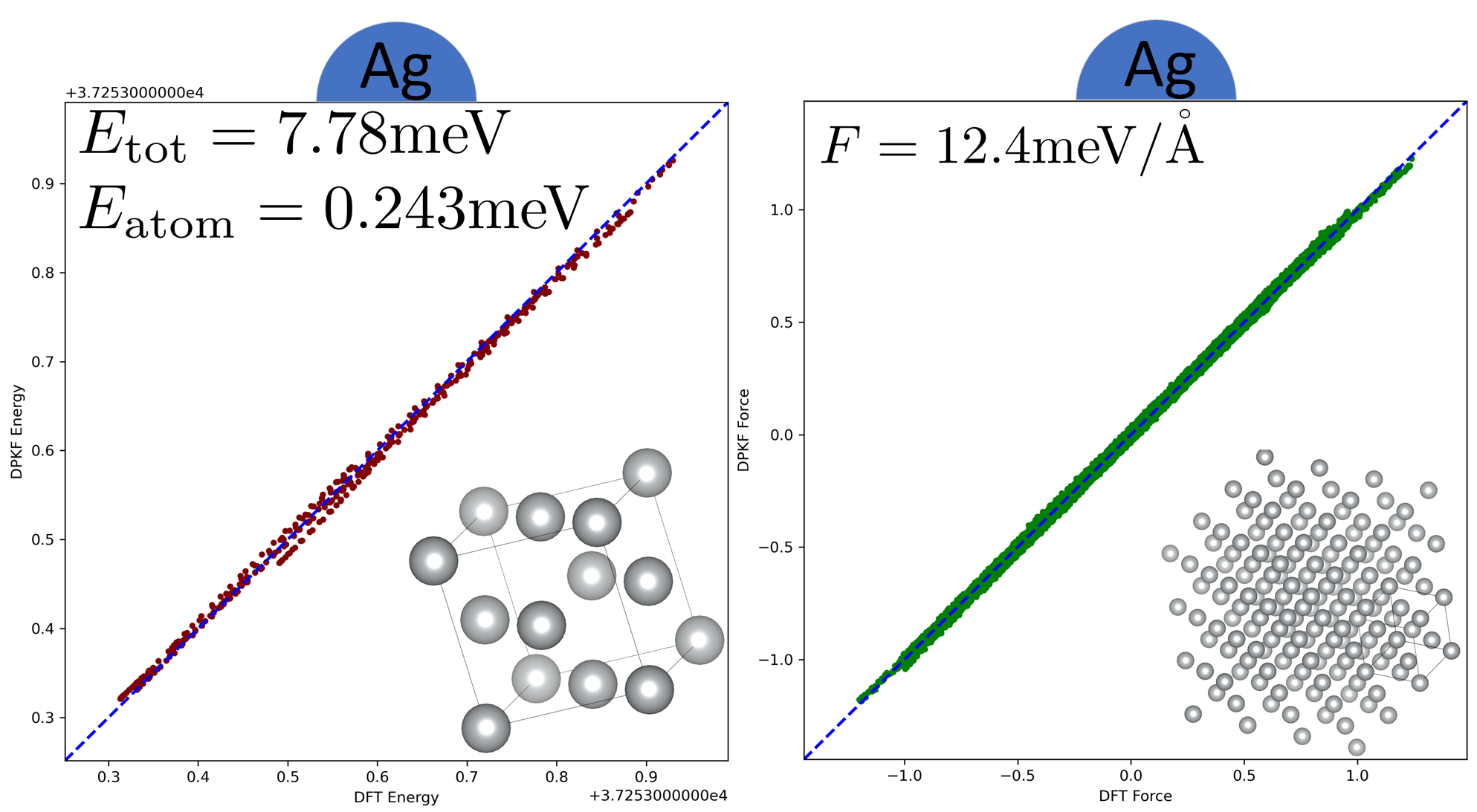} 
		\label{layer[25*25]}}
	\subfigure{
		\includegraphics[width=0.23\textwidth]{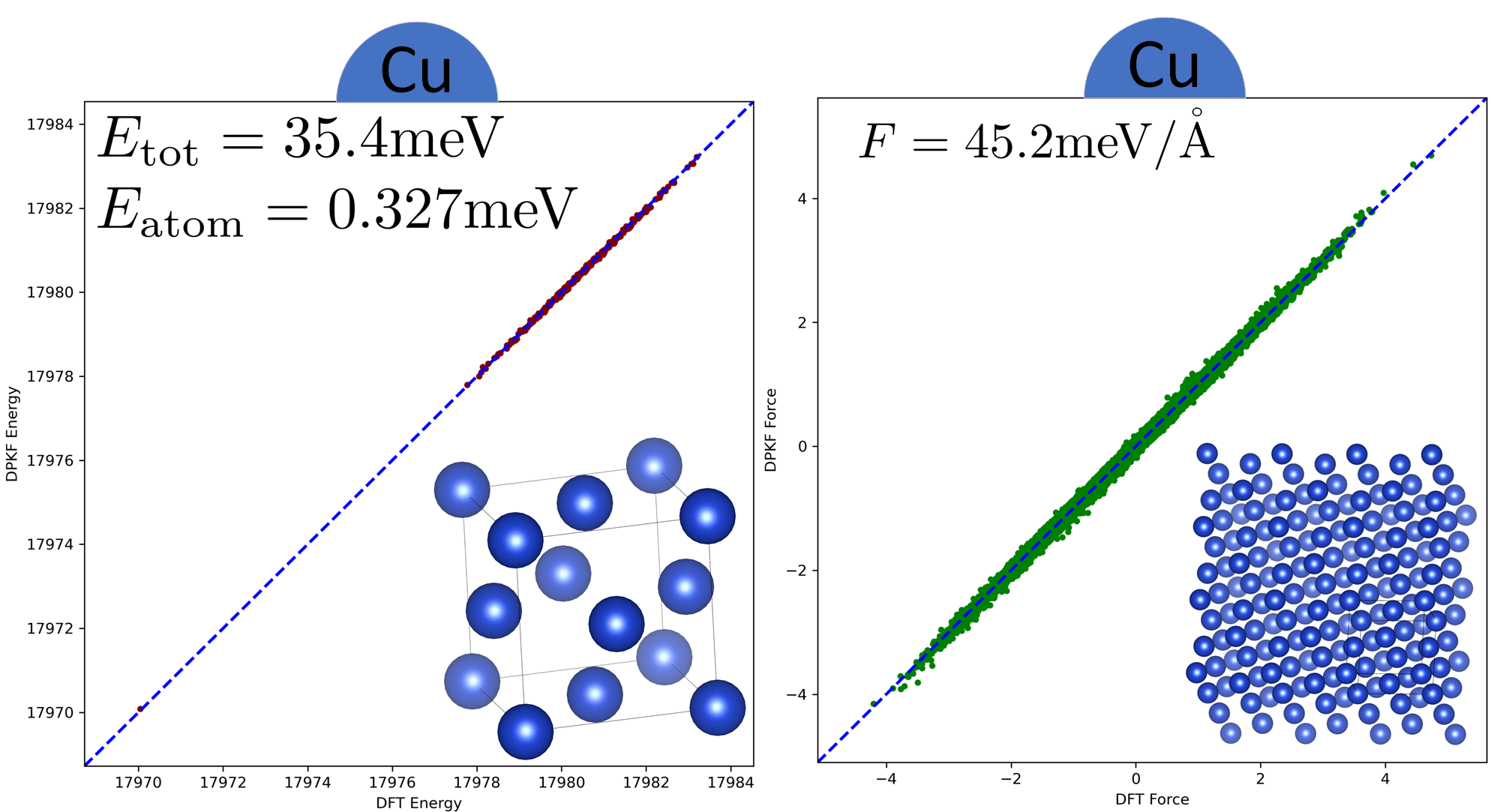} 
		\label{layer[25*25]}}
	\subfigure{
		\includegraphics[width=0.23\textwidth]{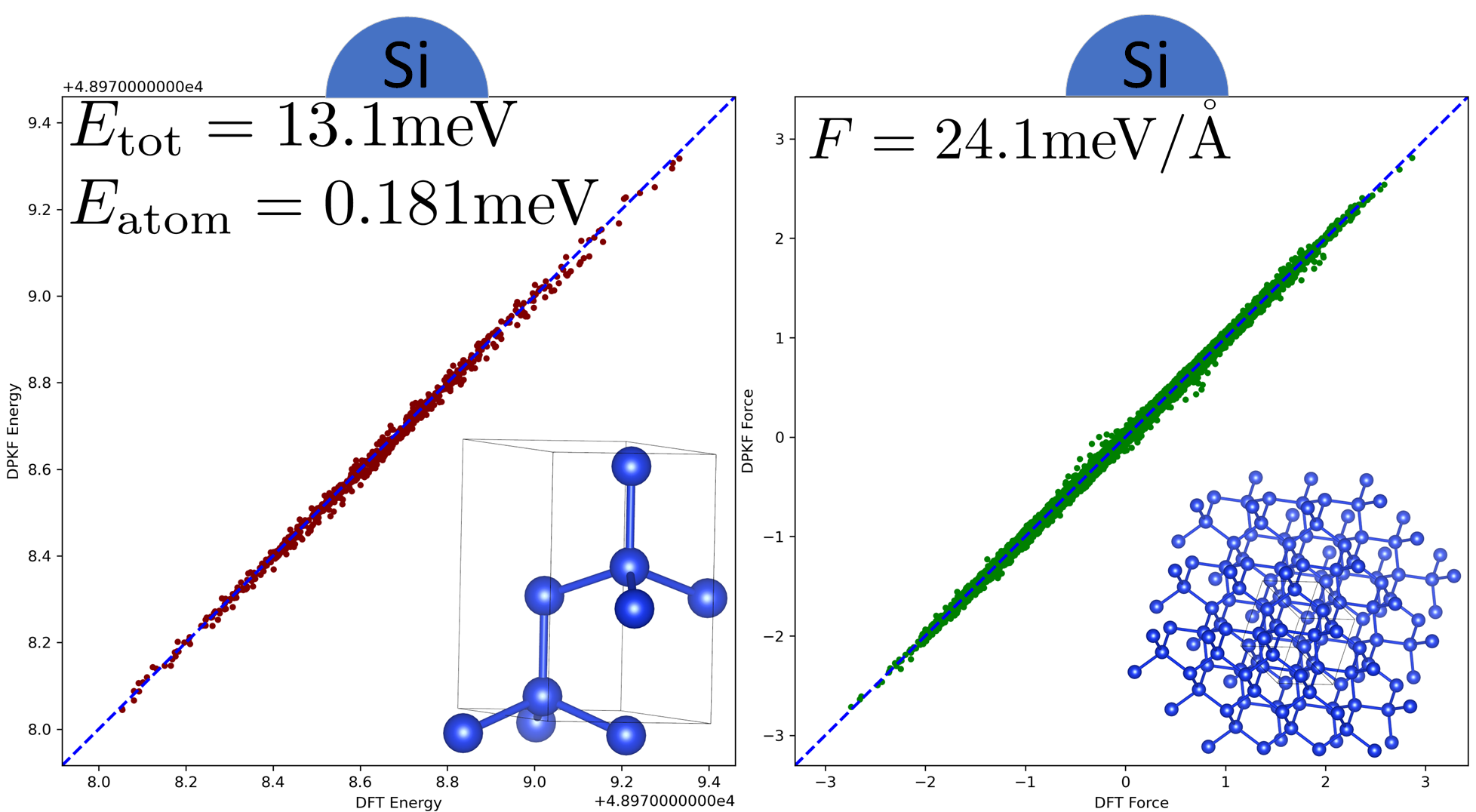} 
		\label{layer[25*25]}}
	\subfigure{
		\includegraphics[width=0.23\textwidth]{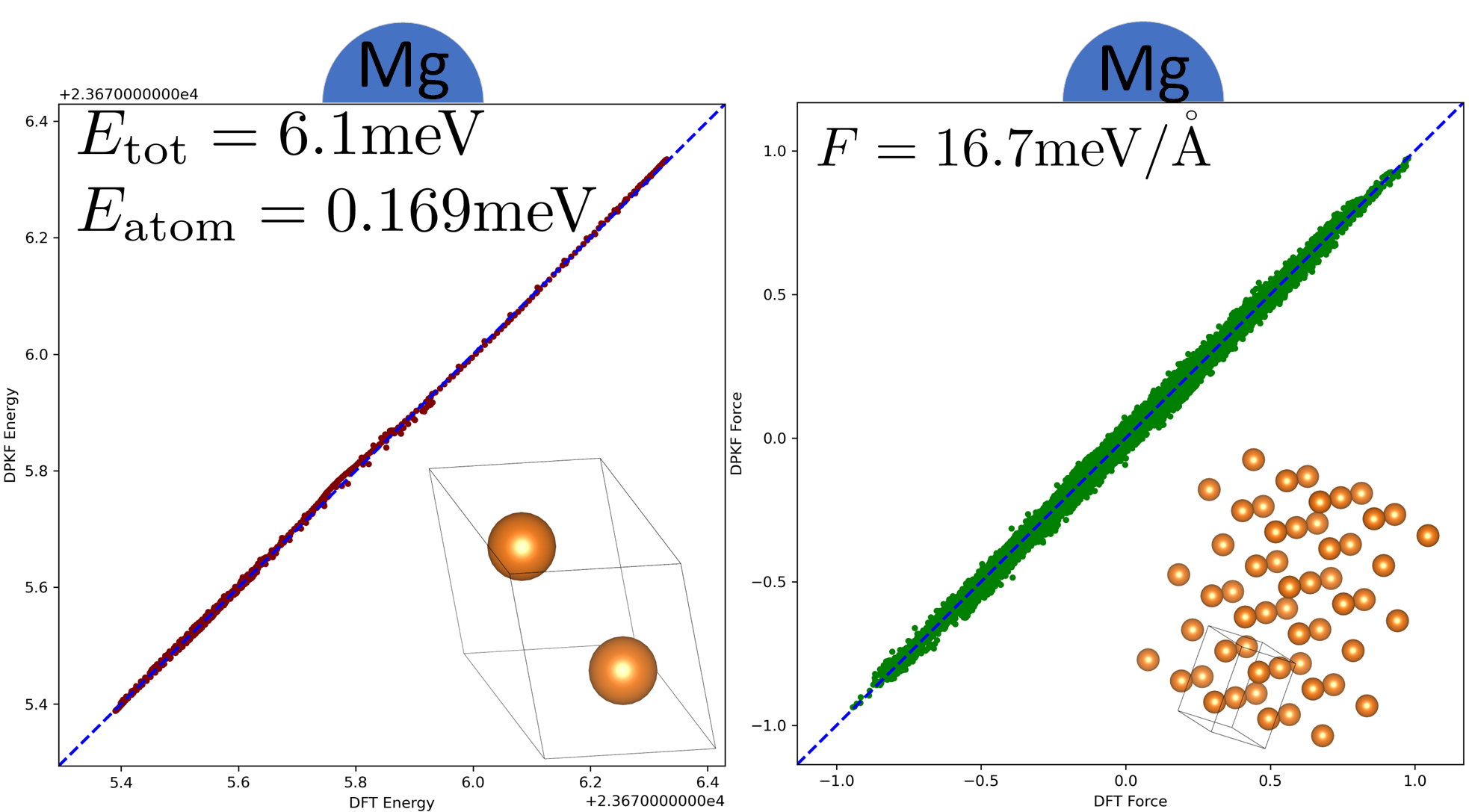} 
		\label{layer[25*25]}}
	\subfigure{
		\includegraphics[width=0.23\textwidth]{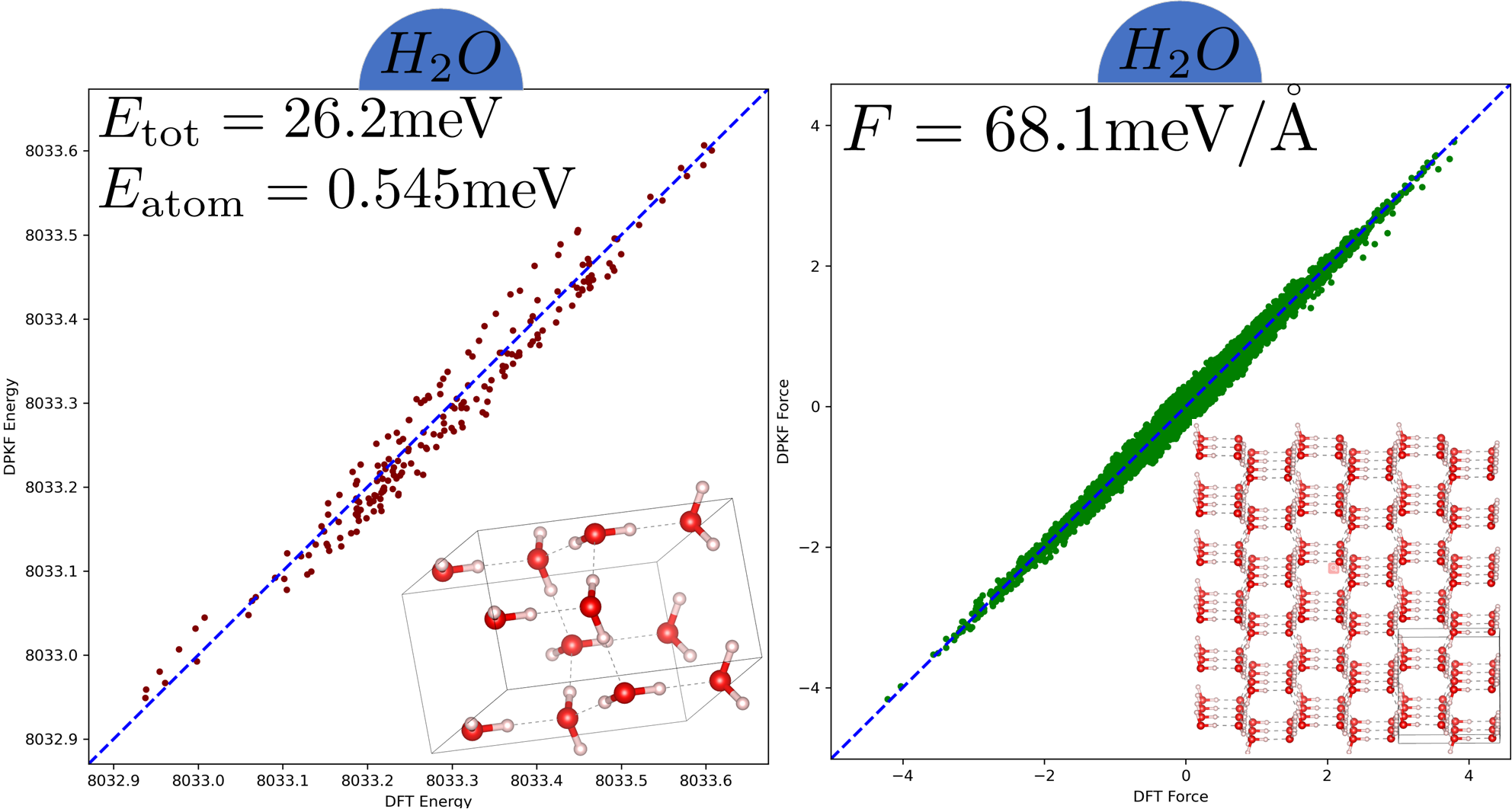} 
		\label{layer[25*25]}}
	\subfigure{
		\includegraphics[width=0.23\textwidth]{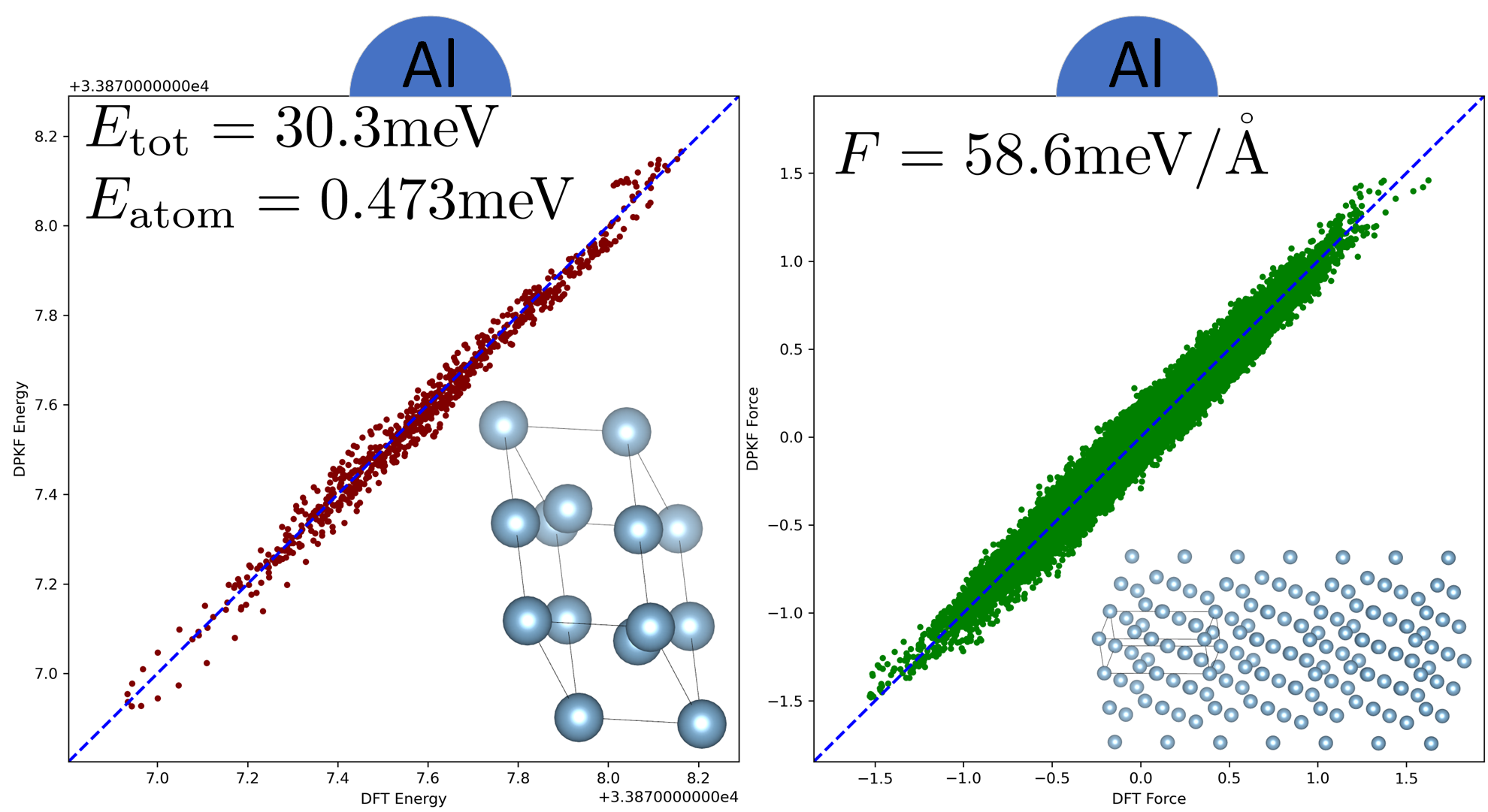} 
		\label{layer[25*25]}}
	\subfigure{
		\includegraphics[width=0.23\textwidth]{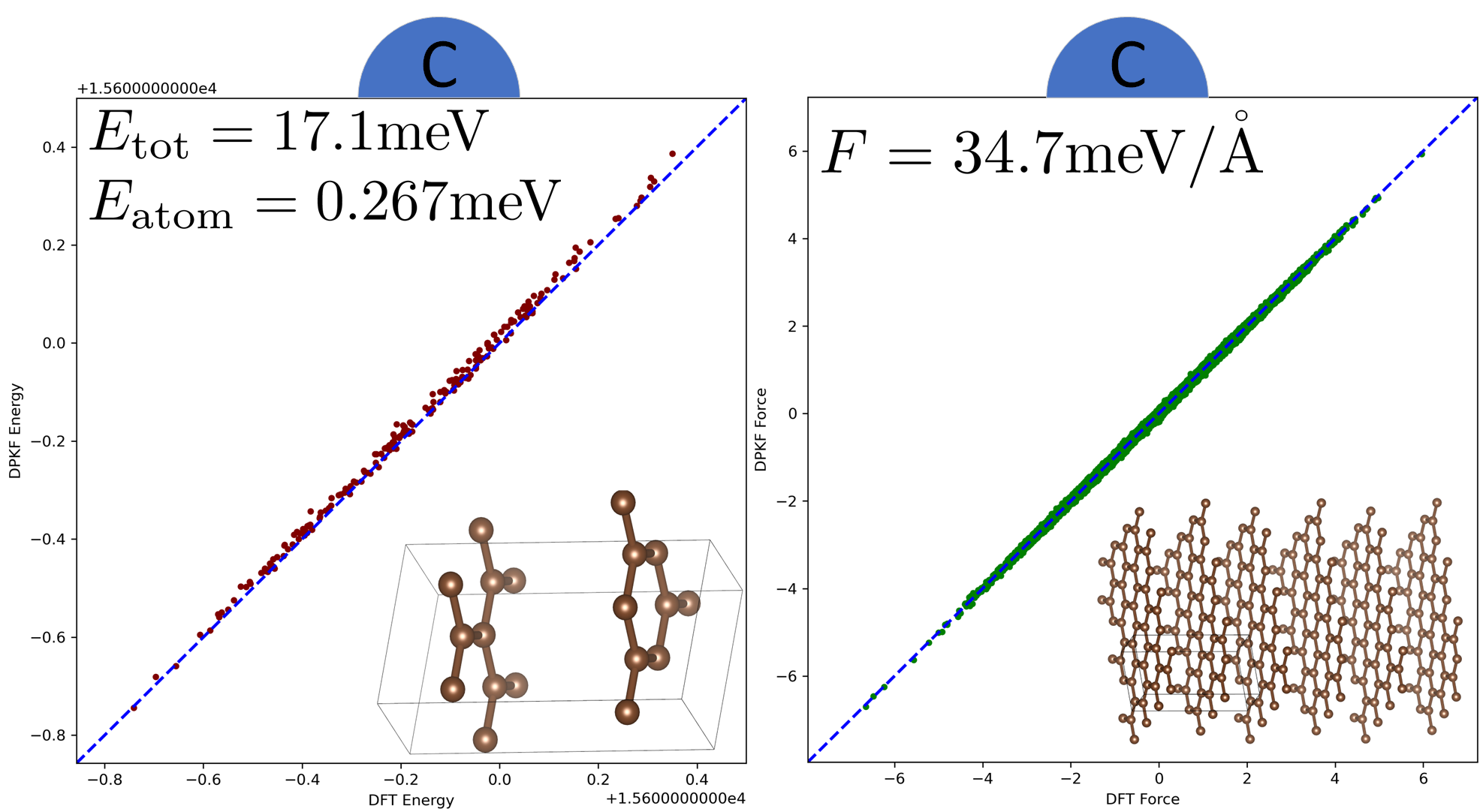} 
		\label{layer[25*25]}}
	\subfigure{
		\includegraphics[width=0.23\textwidth]{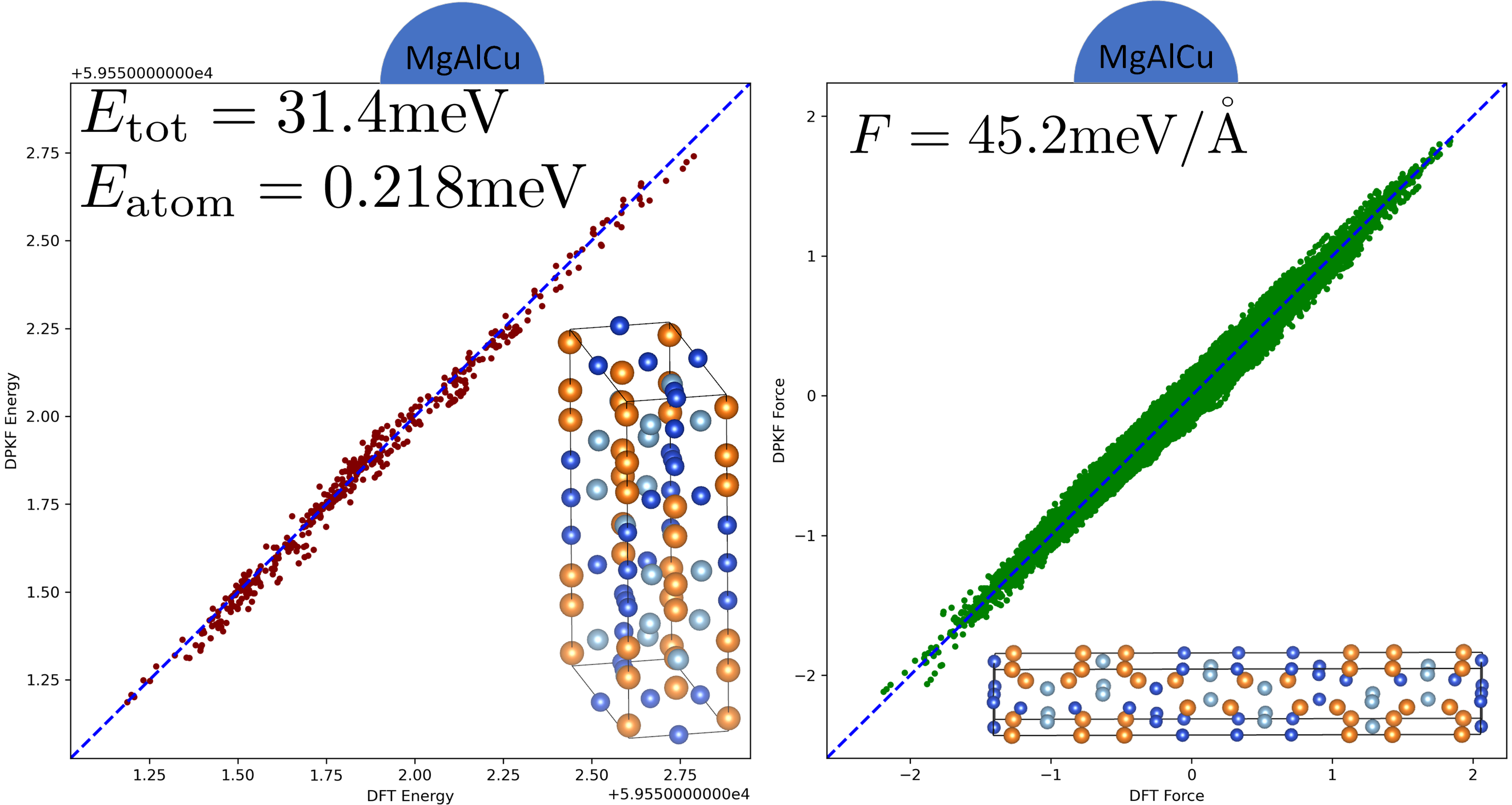} 
		\label{layer[25*25]}}
	\subfigure{
		\includegraphics[width=0.23\textwidth]{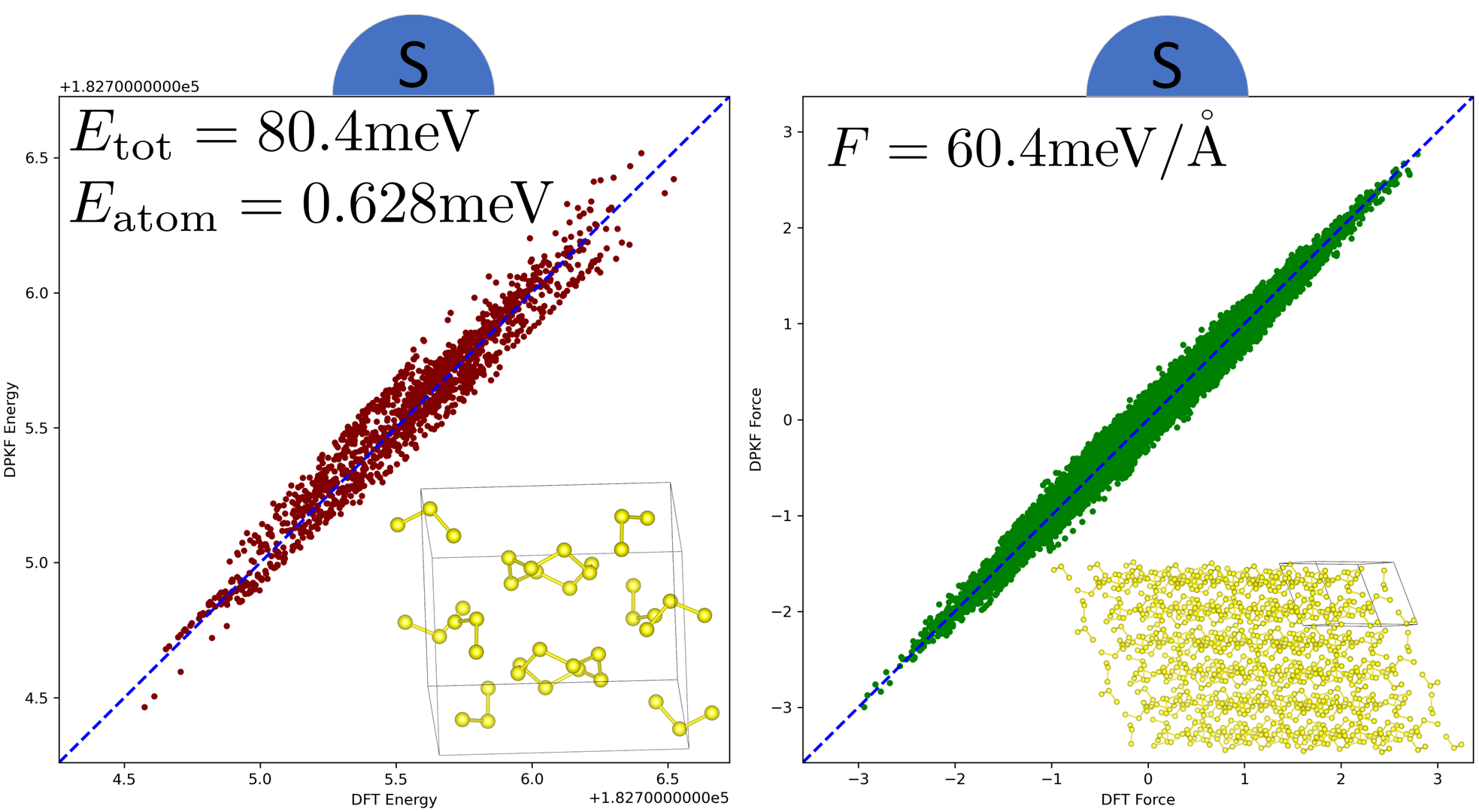} 
		\label{layer[25*25]}}
	\subfigure{
		\includegraphics[width=0.23\textwidth]{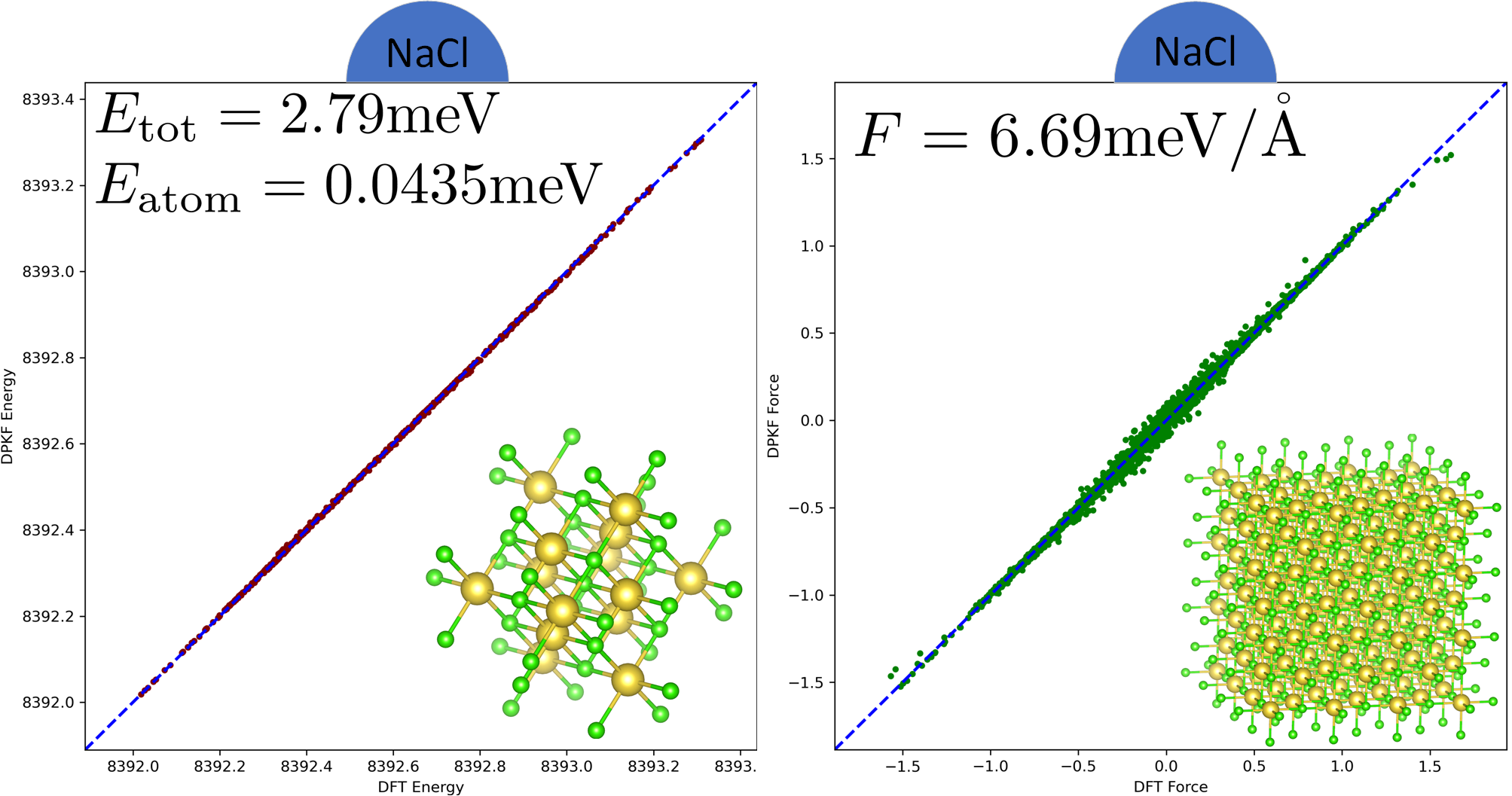} 
		\label{layer[25*25]}}
	\subfigure{
		\includegraphics[width=0.23\textwidth]{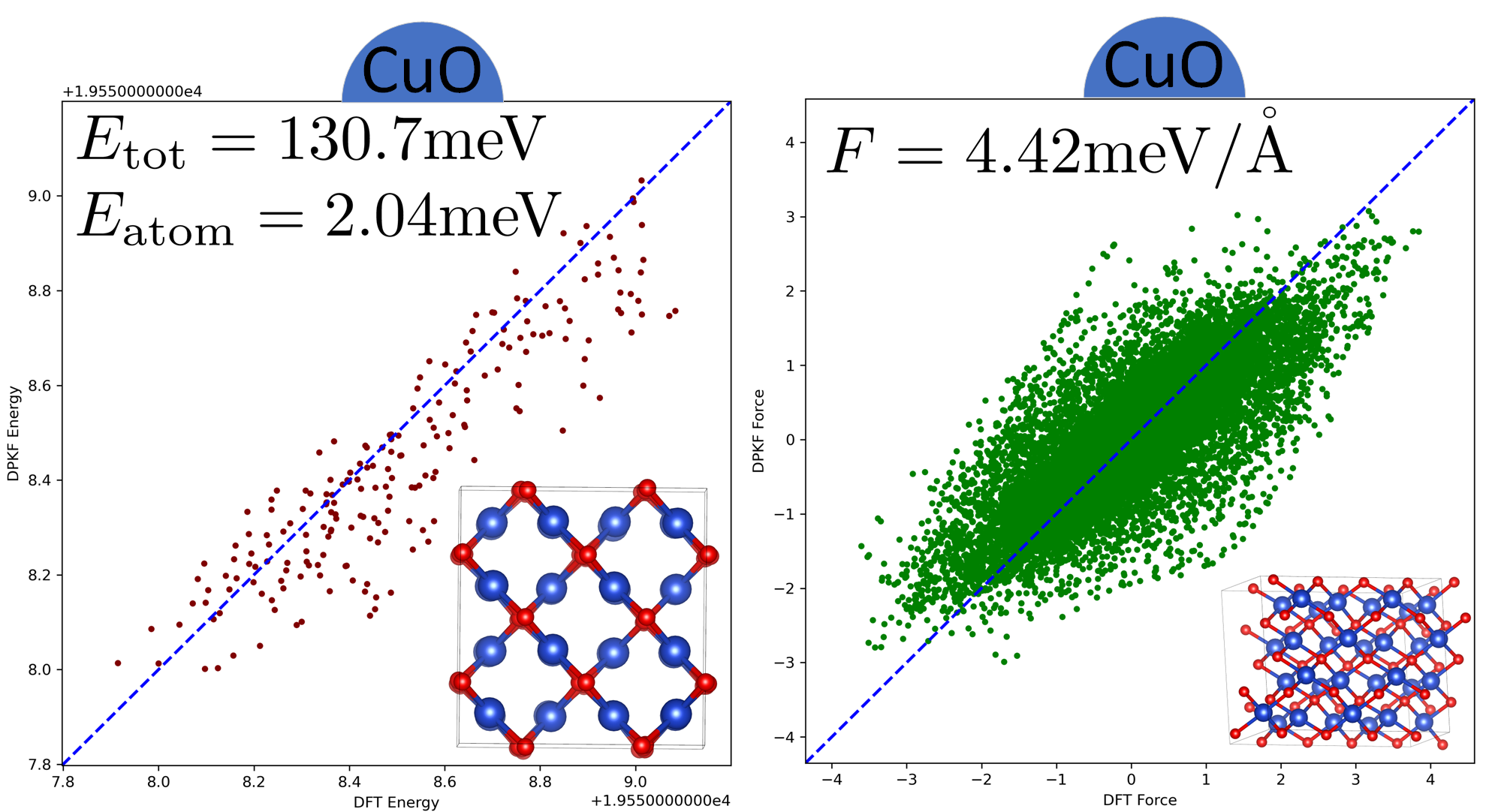} 
		\label{layer[25*25]}}
	\subfigure{
		\includegraphics[width=0.23\textwidth]{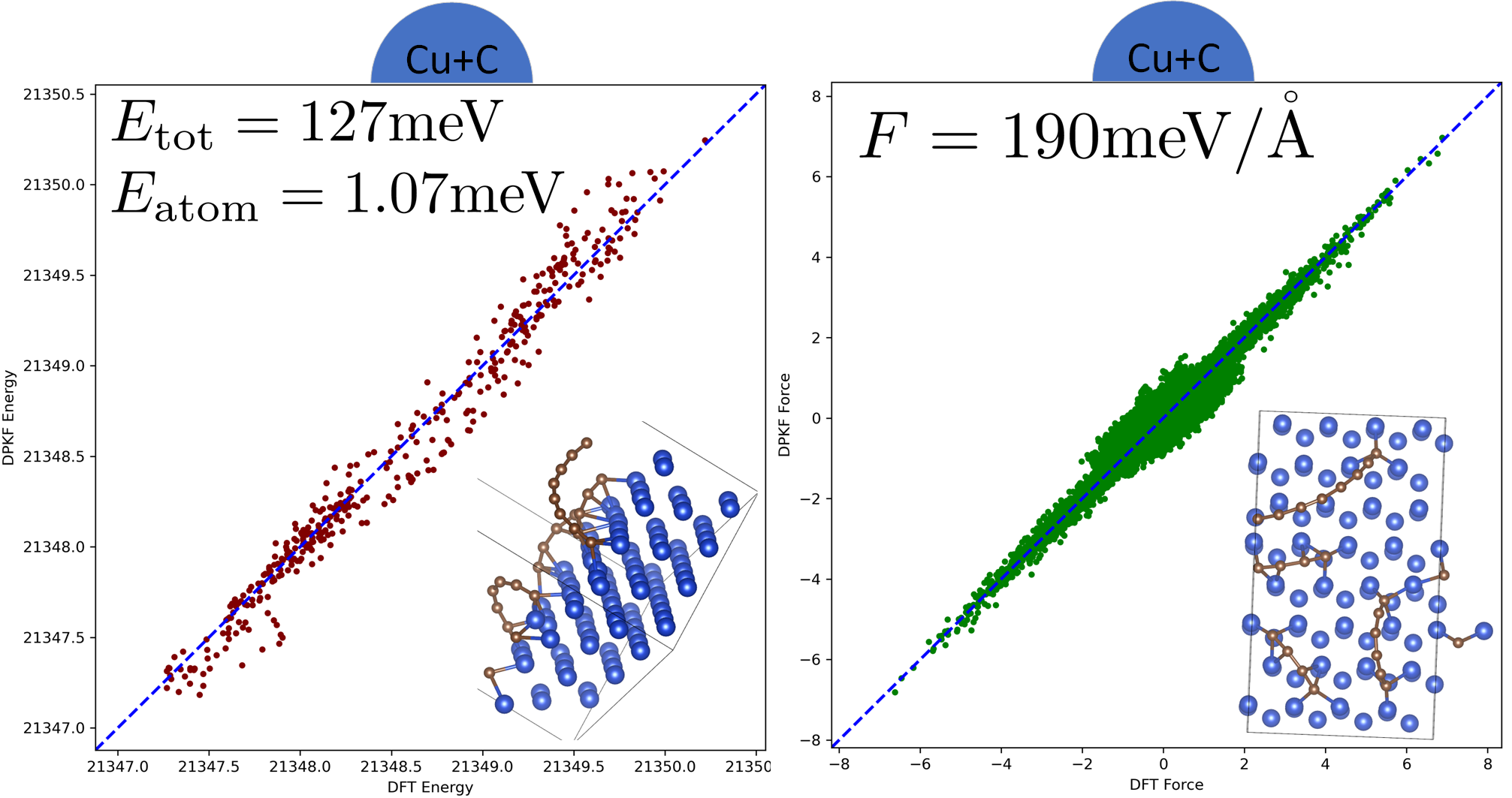} 
		\label{layer[25*25]}}
	\caption{RLEKF performance contrasted with DFT (ground truth) on every snapshot in terms of energy (meV)  and force (meV/$\mathring{\text{A}}$) for various systems. That is to say, the closer the dots approach diagonals, the better the performance is. RMSE of $E_\text{tot}, E_\text{atom}$, and $F$ on the test set, the structure of a unit cell, and extensive structure are shown in the corners of each subfigure.}
	\label{figdiag}
\end{figure*}
\section{Method}
In this section, we introduce RLEKF and its splitting strategy (Fig.~\ref{figovv}) and then compare RLEKF with GEKF.  

\noindent\textbf{The Workflow of Training with RLEKF.} The overview of DP NN with RLEKF is shown in Fig.~\ref{figovv}, and the corresponding training procedures are detailed in Alg.~\ref{alg1}.
Specifically, Alg.~\ref{alg1} and Fig.~\ref{figovv}.b show the training of RLEKF for both energy (Alg.~\ref{alg1}, line~\ref{alg1.1} \& \ref{alg1.2} ) and force (Alg.~\ref{alg1}, line~\ref{alg1.3} \& \ref{alg1.4} ). The forward prediction(shown in Alg.~\ref{alg1} and Fig.~\ref{figovv}.c), backward propagation (shown in Alg.~\ref{alg2} and Fig.~\ref{figovv}.c), and updating of weights based on Kalman filtering with forgetting factor are shown in Alg.~\ref{alg2} (line~\ref{alg2.3} to line~\ref{alg2.4} and Fig.~\ref{figovv}.e). Alg.~\ref{alg1} 
shows the iterative update of $\boldsymbol{w}_{t}$ with $ \mathbf{P}_{t}, \lambda_{t}$ (also shown in Fig.~\ref{figovv}.b). 
After the initialization of $\boldsymbol{w}_{0}$, $\mathbf{P}_{0}$, and $\lambda_{1}$,  the energy process starts and tries to fit the predicted energy to the energy label of a sample, and then the force process updates weights under the supervision of the samples' force label. In the force process for each image, we randomly choose $x$ atoms and concatenate their network-predicted force vectors as an effective predicted force $\hat{F}$, which is then used to update weights. This process (i.e. line~\ref{alg1.3} \& \ref{alg1.4} of Alg.~\ref{alg1}) is repeated $y$ times in a loop. 
Here, we recommend a relatively universal setting $x=6, y=4$. When these two kinds of alternative weights update in the direction of minimizing the trace of $\mathbf{P}_{t}$ for this sample, the same process for the next one repeats until time step $T$. The algorithm RLEKF in line~\ref{alg1.2} \& \ref{alg1.4} of Alg.~\ref{alg1} unfolds below.

\noindent\textbf{RLEKF.}
Alg.~\ref{alg2} shows the layerwise weights-updating strategy of RLEKF, which is an optimized version of EKF in training the DP model. From line~\ref{alg2.1} to line~\ref{alg2.2}, we transform any vector $\hat{Y}$ into a number, which enjoys two advantages. The first is it averages all the updating information in $Y^{\text{DFT}}_{i}-\hat{Y}_{i}$ and transforms a vector into a number which avoids an impregnable problem, solving inversion of overwhelmingly many big matrices introduced later. The second is once Alg.~\ref{alg2} invoked we only need to calculate the gradient for one time. More specifically, from line~\ref{alg2.1} to line~\ref{alg2.2}, we just want to adjust the gradient of $\hat{Y}_{i}$ with respect to weights in the direction of "decreasing the difference between $\hat{Y}_{i}$ and $Y^{\text{DFT}}_{i}$". In order to keep $\mathbf{P}^{out}$ strictly symmetric if  $\mathbf{P}^{in}$ is, the definition of Kalman gain in Alg.~\ref{alg2} is a little different from traditional Kalman filtering. In addition, $\mathbf{a}$ is usually a matrix, but in line~\ref{alg2.3} it degenerates into a number, which heavily reduces the computational complexity, where $L$ is the number of blocks in $\mathbf{P}^{in}$ and take $\alpha_{t}^{2}R _{t}=L\mathbf{I}$ as a suitable hyperparameter. RLEKF independently updates weights and $\mathbf{P}_{t}$ of different layers by KF theory and its splitting strategy introduced in the next subsection. Finally, all updated parameters are \textbf{collect}ed for the prediction of the next turn (Fig.~\ref{figovv}.e, Alg.~\ref{alg2} from line~\ref{alg2.3} to line~\ref{alg2.7}, where $\nu$ is forgetting rate, a hyperparameter describing the varying rate of $\lambda_{t}$ and \textbf{split} is a function for obtaining the $l$-th part after splitting).

\noindent\textbf{The Splitting Strategy of RLEKF.}  According to the splitting strategy of RLEKF, GEKF can be approximated by RLEKF with a reduction in weights error covariance matrix considering the efficiency and stability of a large-scale application. Unlike GEKF, the weights error covariance matrix of whichever time step, up to a scalar factor, $\mathbf{P} =\left\{ P_1,\dots, P_L \right\} $ is a $N \times N$ block diagonal matrix, whose shape is $\left\{ n_{1}\times n_{1}, n_{2}\times n_{2}, \dots, n_L \times n_L \right\}$, where $ n_i$ and $N:=\sum_{i} n_i$ are the number of weights of the $i$th block and that of the whole NN respectively. This means some weights error covariances are forced into $0$ and $\mathbf{P}$ thus is no longer the real weights error covariance matrix at most an approximation, but fortunately there are indeed such good enough approximations that they carry most of the "big values" and information in these diagonal blocks, on which $\mathbf{P}$ concentrates.
As shown in Fig.~\ref{figss}, GEKF concerns correlations between all the parameters. However, it is computationally expensive when adapted to large NNs. As a seemingly good choice, SEKF is a load balanced approximation version of GEKF, but the heavy correlations between parameters in a layer are ignored inappropriately. Overcoming the drawback of SEKF, LEKF can fully consider the internal correlation between parameters in the same layer, whereas unbearable computation still confronts us if the layer contains tens of thousands of parameters and decoupling every small neighboring layer deviates from the load balance idea.
Here, we induce two heuristic principles for choosing a suitable strategy:
\begin{enumerate}
	\item Put weights with high correlation (in the same layer) in the same block if possible. 
	\item The block must not be too large since that burdens computers with much computation (splitting the layers if the number of parameters is larger than the threshold $N_{b}$) or too small since that wastes computational power and loses massive information (gathering the near neighboring small layers). The threshold $N_{b}$ aims at splitting the matrix $\mathbf{P}$ as evenly as possible for load balance and linear computational complexity.
\end{enumerate}
Therefore, following these principles, we induce the splitting strategy of RLEKF (Fig.~\ref{figovv}.d) and reorganize the weights in different layers, named parameter parts, into several more appropriate layers. Trainable parameters could be decomposed into a series of $l$ the parts of size $p_{1}, p_{2}, \dots, p_{l}$. Starting from the first part of size $p_{1}$, we execute the commands below repeatedly until $z=l$:
\begin{enumerate}
	\item For a given part with $p_{s}$ weights, we split them into $\lfloor\frac{p_{s}}{N_{b}}\rfloor$ layers of size $N_{b}$ and a new part of size $p_{s}\leftarrow p_{s}-N_{b}\lfloor\frac{p_{s}}{N_{b}}\rfloor$. 
	\item Then consider subsequent parts with $p_{s}, p_{s+1}, \dots, p_{z}$ weights, gather them together into a layer of size $\sum_{i=s}^{z}p_{i}$ and then set $s\leftarrow z+1$ , if $\sum_{i=s}^{z}p_{i}\leq N_{b}$ and $(\sum_{i=s}^{z+1}p_{i}>N_{b}$ or $z=l)$.
\end{enumerate}
Empirically, Fig.~\ref{figss} validates the efficiency of RLEKF, comparably or even more accurate than GEKF.

\section{Theoretical Analysis}
In this section, we prove the convergence of weights updating and the stability of the training process (avoid gradient exploding). For simplicity, we analyze GEFK case which has the same asymptotic feature as RLEKF~\cite{doi:10.1021/ct049976i}. 
\begin{thm}
	In EKF problem \eqref{eq:ekf}, setting $\alpha_{t}^{2}R _{t}=L\mathbf{I}$, assuming components of $\mathbf {H} _{i}$ are independent and subject to identical distribution with mean 0 and variance $\sigma^{2}$, we have
	\begin{equation*}
		\epsilon_{t}\mathbf {K}_{t}\sim\mathcal{O}(\frac{1}{t})
	\end{equation*}
	with the probability arbitrarily close to 1 when $t\to \infty$, which means the convergence of weights updating and thus the algorithm avoidance of gradient exploding.
	\label{thm:main}
\end{thm}
The following is a brief proof of Theorem \ref{thm:main}. Based on basic KF theory, we obtain
\begin{align*}
	\mathbf{P}_{t}&=\lambda _{t}^{-1}\mathbf {P} _{t-1}-\lambda _{t}^{-2}\mathbf {P} _{t-1}\mathbf {H} _{t}^{\textsf {T}}\mathbf{a} _{t}^{-1}\mathbf {H} _{t}\mathbf {P} _{t-1}L^{-1},\\
	\mathbf{a}_{t}&=\lambda _{t}^{-1}\mathbf {H}^{\textsf {T}} _{t}\mathbf {P} _{t-1}\mathbf {H} _{t}L^{-1}+1=\mathbb{E}[\epsilon_{t}^{2}]\alpha_{t}^{2},\\
	\lambda_{t}&=1-(1-\lambda_{1})\nu^{t-1},\\
	\mathbf{P}_{t}&=\mathbb{E}[\tilde{{\boldsymbol{w}}}_{t}\tilde{{\boldsymbol{w}}}_{t}^{\textsf{T}}]\alpha_{t}^{2},
\end{align*}
and 
\begin{equation*}
	\mathbf {P}_{t}^{-1}=\lambda _{t}\mathbf {P}^{-1}_{t-1}+\mathbf {H} _{t}\mathbf{H} _{t}^{\textsf {T}}L^{-1}
\end{equation*}
by using Woodbury matrix identity, where $\tilde{{\boldsymbol{w}}}_{t}=\boldsymbol{w}-\boldsymbol{w}_{t}$, weights error covariance matrix $\mathbb{E}[\tilde{{\boldsymbol{w}}}_{t}\tilde{{\boldsymbol{w}}}_{t}^{\textsf{T}}]=(\mathbf {P}_{0}^{-1}+\sum_{i=1}^{t}\alpha_{i}^{2}\mathbf {H} _{i}\mathbf{H} _{i}^{\textsf {T}}L^{-1})^{-1}$.
In the following experiments, we assume components of $\mathbf {H} _{i}$ are independent and subject to identical distribution with mean 0 and variance $\sigma^{2}$. So, the covariance matrix of $\mathbf {H} _{i}$ is $\sigma^{2}\mathbf{I}$ and $\mathbf {P}_{0}=\mathbf{I}$ . Hence
\begin{equation*}
	\mathbb{E}[\mathbf {P}_{t}^{-1}]=\mathbf{I}+\sum_{k=1}^{t}\alpha_{k}^{2}\alpha_{t}^{-2}L^{-1}\sigma^{2}\mathbf{I}.
\end{equation*}
Using $q$-Pochhammer symbol, we find
\begin{equation*}
	\lim_{t\to\infty}\alpha_{t}^{2}=\Pi_{i=0}^{\infty}(1-(1-\lambda_{1})\nu^{i})=((1-\lambda_{1});\nu)_{\infty}=\alpha
\end{equation*}
exists. Therefore, 
$S(t):=\sum_{k=1}^{t}\alpha_{k}^{2}\alpha_{t}^{-2}$ of order $\mathcal{O}(t)$.
According to the law of large numbers, we get
\begin{equation*}
	\lim_{t\to\infty}\frac{\mathbf {P}_{t}^{-1}}{S(t)}\xrightarrow{a.s.}\sigma^{2}L^{-1}\mathbf{I},
\end{equation*}
i.e.
\begin{equation*}
	\lim_{t\to\infty}\mathbf {P}_{t}\stackrel{a.s.}{=\joinrel=\joinrel=}\lim_{t\to\infty} \frac{L\mathbf{I}}{\sigma^{2}S(t)} .
\end{equation*}
Hence,
\begin{equation*}
	\mathbf {K}_{t}\stackrel{a.s.}\sim\mathcal{O}(\frac{1}{t}),
\end{equation*}
if $t$ is large enough. Further, $\epsilon_{t}$ is bounded with a probability 
\begin{equation*}
	\mathbb{P}(|\epsilon_{t}|\leq B)\geq 1-\frac{\mathbb{E}[\epsilon_{t}^{2}]}{B^{2}}=1-\frac{\alpha_{t}^{-2}(\lambda _{t}^{-1}\mathbf {H}^{\textsf {T}} _{t}\mathbf {P} _{t-1}\mathbf {H} _{t}+1)}{B^{2}},
\end{equation*}
arbitrarily close to 1, where Markov's inequality is used, if initialization $\boldsymbol{w}_{0}$ is close enough to some local minimum $\boldsymbol{w}^{*}$ of the landscape for Taylor approximation \eqref{eq:Taylor approximation}. Therefore, 
\begin{equation*}
	\epsilon_{t}\mathbf {K}_{t}\sim\mathcal{O}(\frac{1}{t})
\end{equation*}
with the probability arbitrarily close to 1.

\begin{figure}[t]
	\centering
	\includegraphics[width=0.44\textwidth]{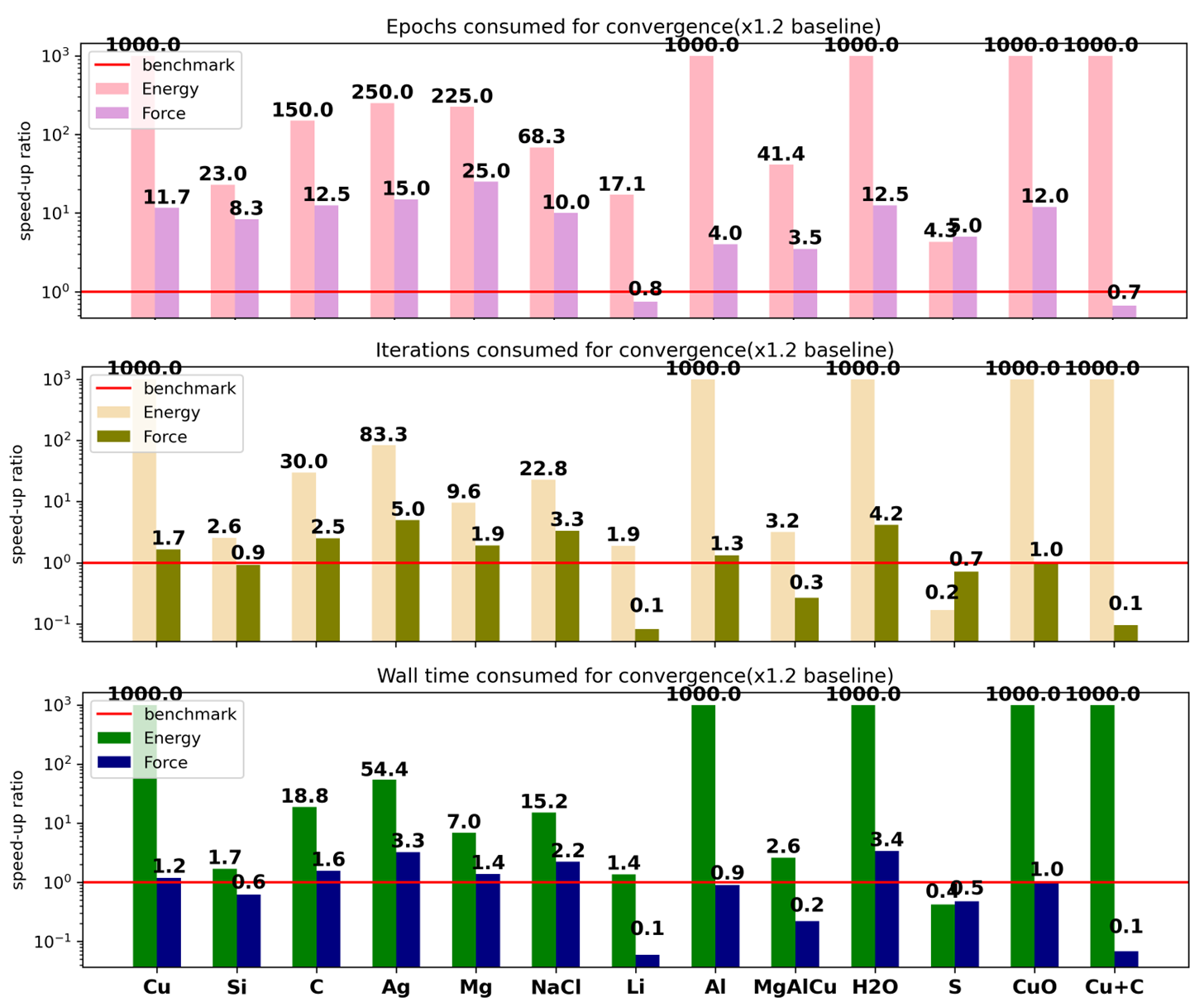} 
	\caption{Speed ratio of RLEKF's reaching an accuracy baseline to Adam's in terms of Energy and Force according to epoch, iteration(weights updating) and, wall-clock time, where 1000 means Adam can not reach the baseline. The baseline is 1.2$\times$ the lower RMSE between Adam and RLEKF on the test datasets in Tab.~\ref{tab:rmse}.}
	\label{figspd}
\end{figure}
\begin{figure}[t]
	\centering
	\subfigure{
		\includegraphics[width=0.22\textwidth]{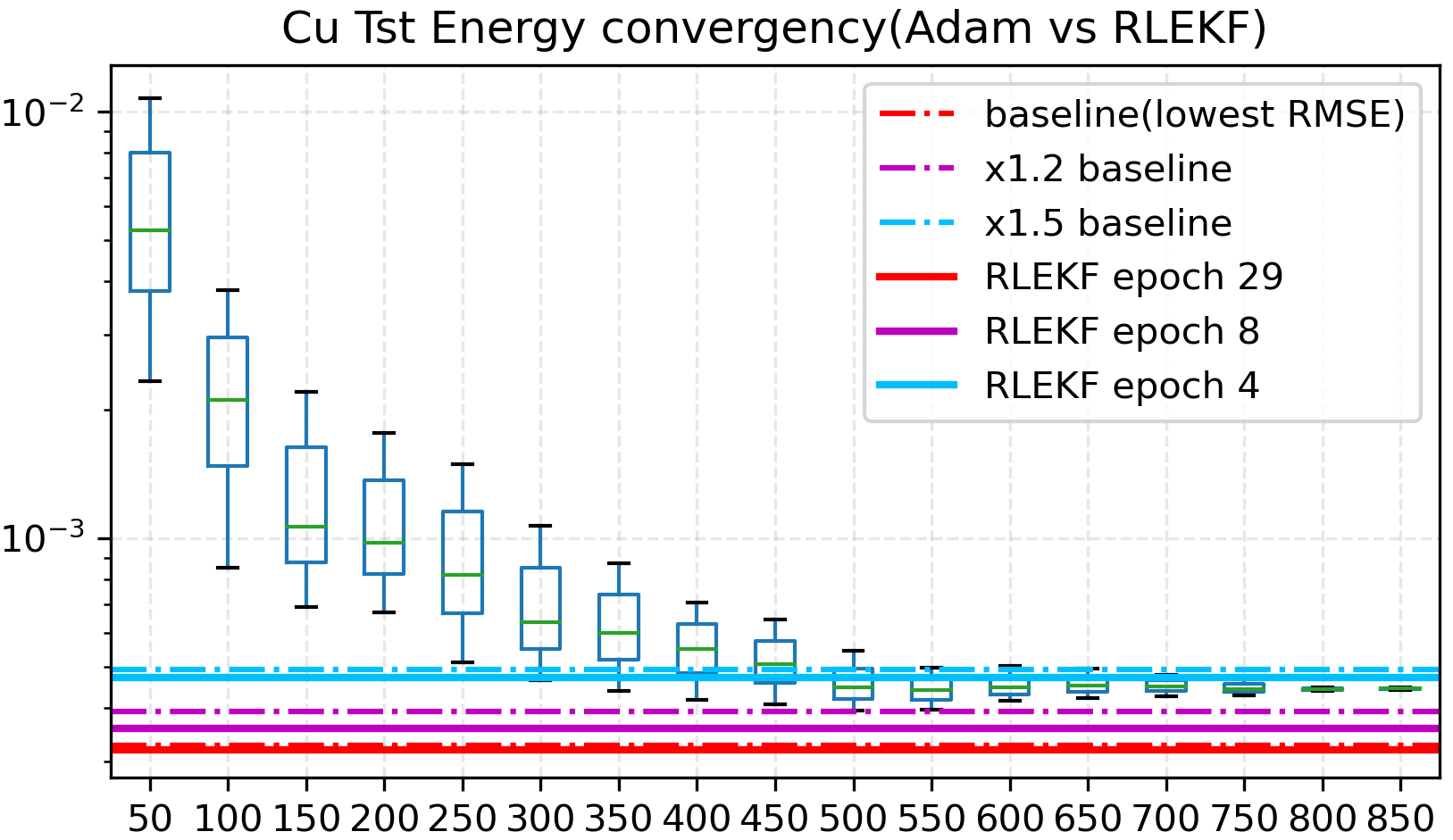} 
		\label{layer[25*25]}}
	\subfigure{
		\includegraphics[width=0.22\textwidth]{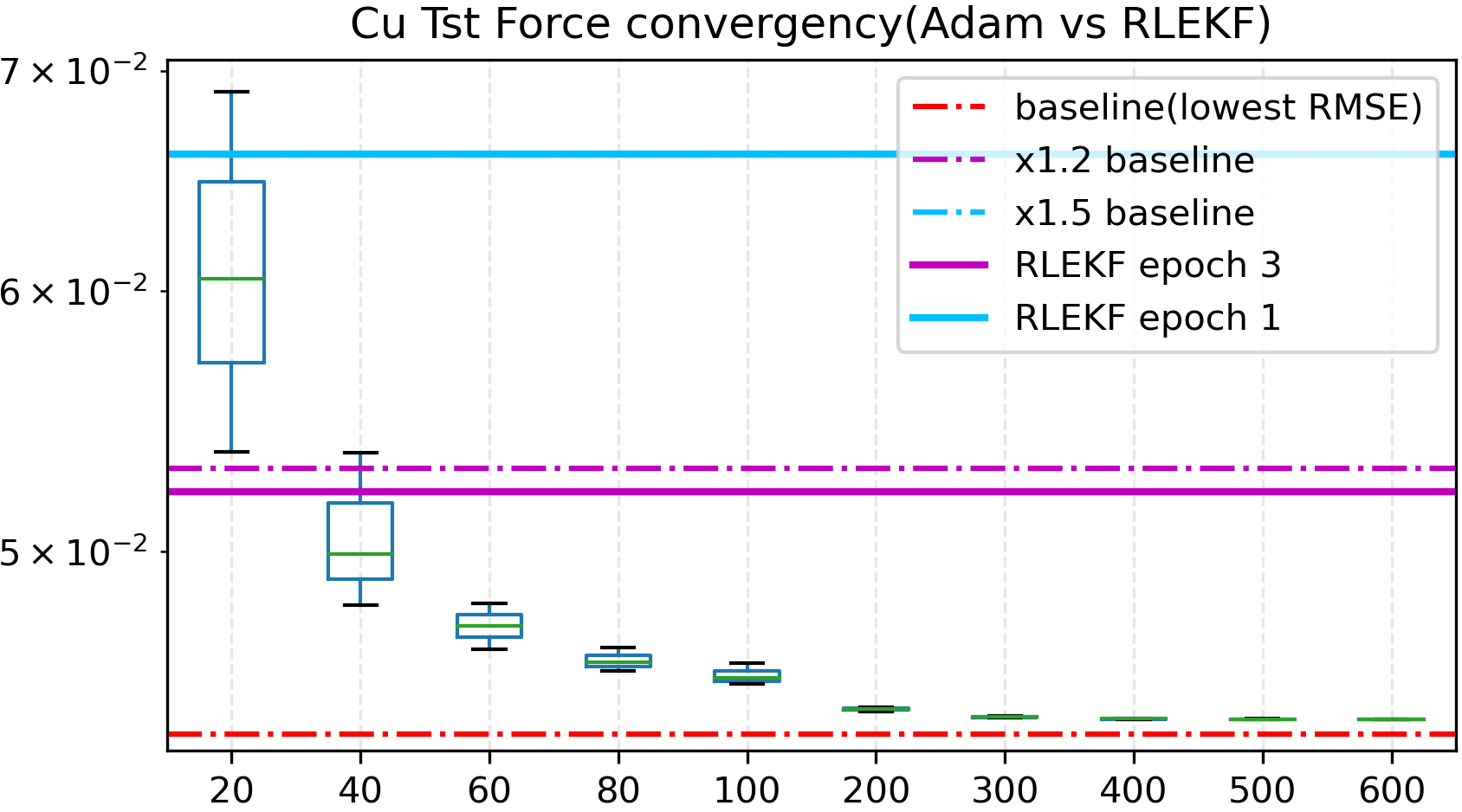} 
		\label{layer[25*1]} }
	\caption{Convergence of RMSE of bulk copper. The boxplots show the relationship between the test RMSE and the training epoch of Adam. For the energy case, each box on $x$ stands for the statistics (median, the first quaritle, and the third quartile) on RMSE data between epoch $(x-50)$ and epoch $(x+50)$ (replacing 50 with 20 for force). The RMSEs of RLEKF after several epochs are shown in lines. The remaining systems refer to supplementary.}
	\label{figcvg}
\end{figure}
\begin{figure}[t]
	\centering
	\includegraphics[width=0.41\textwidth]{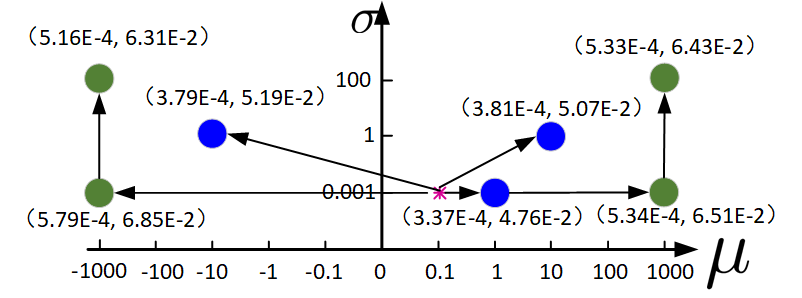}
	\caption{Accuracy (energy and force RMSE on a test dataset) of Cu models with different skip connection weights initialization $\mathcal{N}(\mu, \sigma^2)$. The red star represents the setup adopted in the above experiments, to whose result Fig.~\ref{tab:rmse} refers. The 3 blue dots and the red star show similar accuracy, the 4 green dots show lower precision but are still very reasonable.}
	\label{figrbst}
\end{figure}
\section{Experiments and Results}
Bulk systems are challenging AIMD tasks due to their extensiveness (periodic boundary condition) and complexity (many different phases and atomic components). Our experiment is conducted on several representative problematic bulk systems. They are simple metal (Cu, Ag, Mg, Al, Li), alloy (MgAlCu), nonmetal (S, C), semiconductor(Si), simple compound (H$_{2}$O), electrolyte (NaCl), and some challenging systems (CuO, Cu$+$C). The effectiveness of RLEKF is shown by a comparison with the SOTA benchmark Adam in terms of the RMSEs of predicted total energy of the whole system $E_{\text{tot}}$, and forces $ \left\{ (F_{x,i}, F_{y,i}, F_{z,i}) | i=1, \dots, n \right\} $, where $x,y,z$ correspond to different directions of Cartesian coordinate system and $i$ is the index of atom (Tab.~\ref{tab:rmse}). 

\noindent\textbf{Experiment Setting.} Set $\lambda_{1}=0.98$, $\mathbf {P}_{0}=\mathbf{I}$, $\nu=0.9987$, $\boldsymbol{w}_{0}$ consistent with DeePMD-kit~\cite{wang2018deepmd}. The network configuration is [1, 25, 25, 25] (embedding net), [400, 50, 50, 50,1] (fitting net).

\noindent\textbf{Data Description.} We choose 13 unbiased and representative datasets of the aforementioned systems with certain specific structures (second column of Tab.~\ref{tab:rmse}). For each dataset, snapshots are yielded based on solving \textit{ab initio} molecular trajectories via PWmat~\cite{JIA2013102}. During this process, to enlarge the sampling span of configuration space, we fast generate a long sequence of the snapshot by small time step (the third column of Tab.~\ref{tab:rmse}) and choose one for every fixed number in the temperature 300K. 

\noindent\textbf{Main Results.} Both RLEKF and Adam yield good results except for CuO and Cu$+$C systems (Tab.~\ref{tab:rmse}). From an \textbf{accuracy} perspective, RLEKF reaches a higher energy accuracy in 12 cases except for Si, little less accurate than Adam. As for force, RLEKF is more accurate than Adam in 7 cases while the reminder achieves a very comparable precision. Furthermore, Fig.~\ref{figdiag} shows how far the fitting results of RLEKF deviate from those of DFT (ground truth). From a \textbf{speed} perspective, generally, RLEKF converges to a reasonable RMSE much faster than Adam in most of the 13 cases in terms of energy (Fig.~\ref{figspd}). We take bulk copper as an example to demonstrate how to understand information in Fig.~\ref{figspd}. The lower Energy and Force RMSE between Adam and RLEKF is 0.327 meV and 44.4 meV/$\mathring{A}$. 
Therefore, the \textbf{1.2$\times$ energy baseline} is 0.327$\times$1.2=0.392 meV. It is unreachable for Adam, which is denoted as 1000 in (Fig.~\ref{figcvg}). The \textbf{1.2$\times$ force baseline} is 44.4$\times$1.2=52.6meV/$\mathring{\text{A}}$. Adam spends 35 epochs to reach the baseline while RLEKF's only costs 3 epochs (Fig.~\ref{figcvg}). Then, the force speed ratio is 35/3=11.67 according to epoch. For each sample, Adam updates weights for 1 time and RLEKF updates for 7 times (1 time in the energy process and 6 times in the force process). The iteration speed ratio is 35/(3$\times$7)=1.66. On tesla V100, the time cost of each epoch is 60s (Adam) and 587s (RLEKF). Thus, the wall-clock time speed ratio is 35$\times$60/(3$\times$587)=1.19.

\noindent\textbf{Robustness Analysis: Distribution of Hyperparameter of Weights Initialization.} RLEKF is very stable as an optimizer, which can keep NNs from gradients exploding and consequently endow them with very loose initialization constraints almost up to none as shown in Fig.~\ref{figrbst}.

\noindent\textbf{Computational Complexity.} RLEKF also benefits from computing through reducing the computation compared to GEKF. There are 3 computational intensive parts in Alg.~\ref{alg2}, calculating $a$ (line~\ref{alg2.3}), $K$ (line~\ref{alg2.5}), and $P^{out}$ (line~\ref{alg2.6}). Due to the even splitting strategy, the order of float operation for each block is $\mathcal{O}(N_{b}^{2})$, and the number of the block is of order $ \mathcal{O}(N/N_{b})$. Hence the total computational complexity of RLEKF is of order $\mathcal{O}((N/N_{b}) N_{b}^{2})=\mathcal{O}(NN_{b})$.

\section{Conclusions}
We proposed an optimizer RLEKF on DP NN and tested RLEKF on several typical bulk systems (simple metals, insulators, and semiconductors) of diverse structure (FCC, BCC, HCP). RLEKF defeated SOTA benchmark Adam by 11-1 (7-6) in precision and 12-1 (7-6) in wall-clock time speed in terms of energy (force). Besides, the convergence of weights updating is proved theoretically and RLEKF presents robustness on weights initialization. 
To sum up, RLEKF is an accurate, efficient, and stable optimizer, which paves another path to training general large NNs.

\section{Acknowledgments}

This work is supported by the following funding: National Key Research and Development Program of China (2021YFB0300600), National Science Foundation of China (T2125013, 62032023, 61972377), CAS Project for Young Scientists in Basic Research (YSBR-005) and Network Information Project of Chinese Academy of Sciences (CASWX2021SF-0103), the Key Research Program of the Chinese Academy of Sciences grant No. ZDBS-SSW-WHC002, Soft Science Research Project of Guangdong Province (No. 2017B030301013), and Huawei Technologies Co., Ltd.. We thank Dr. Haibo Li for helpful discussions.

\bibliography{FKF}
\end{document}